\documentclass[usenatbib]{mnras}
\usepackage{aas_macros}
\usepackage{amsmath}
\usepackage{amssymb}
\usepackage{color}
\usepackage{epsfig}
\usepackage{float}
\usepackage{graphicx}
\usepackage{latexsym}
\usepackage{morefloats}
\usepackage{natbib}
\usepackage{subfigure}
\usepackage{times}
\newcommand{\plotone}[1]{\resizebox{0.95\hsize}{!}{\includegraphics{#1}}}

\newcommand{\plottwo}[2]{\resizebox{0.95\hsize}{!}
{\includegraphics{#1}\hspace{.5cm}\includegraphics{#2}}}
\newcommand{\plottwotwo}[2]{\resizebox{0.85\hsize}{!}
{\includegraphics{#1}\hspace{.5cm}\includegraphics{#2}}}

\newcommand{\plotx}[1]{\resizebox{0.75\hsize}{!}{\includegraphics{#1}}}

\newcommand{\Gyr}{\ensuremath{~\rm Gyr}}

\newcommand{\Msun}{{~\rm M_\odot}}

\newcommand{\kpc}{~\rm kpc}
\newcommand{\Mpc}{~\rm Mpc}

\newcommand{\uvec}[1]{\boldsymbol{\mathit{\hat{#1}}}}

\def\gsim { \lower .75ex \hbox{$\sim$} \llap{\raise .27ex \hbox{$>$}}}
\def\lsim { \lower .75ex \hbox{$\sim$} \llap{\raise .27ex \hbox{$<$}}}
\newcommand{\eagle}{\textsc{eagle}}

\newcommand{\simRef}{Ref-L{\small 0100}N{\small 1504}}
\newcommand{\lcdm}{$\Lambda$CDM}
\newcommand{\refeq}[1]{Eq. \eqref{#1}}
\newcommand{\refsec}[1]{Section \ref{#1}}
\newcommand{\reftab}[1]{Table \ref{#1}}
\newcommand{\reffig}[1]{Fig. \ref{#1}}

\newcommand{\MWthin}{\emph{MW-like-thin}}
\newcommand{\MWorbit}{\emph{MW-like-orbit}}

\title[Evolution of planes of satellites]
{Evolution of galactic planes of satellites in the \textsc{eagle} simulation}
\author[Shao et al.]
{\parbox{\textwidth}{
 Shi Shao,$^{1}$\thanks{E-mail: shi.shao@durham.ac.uk} Marius Cautun$^{1}$ and
 Carlos S. Frenk$^{1}$ \vspace{.20cm}}\\
$^1$Institute for Computational Cosmology, Department of Physics, Durham University, South Road Durham DH1 3LE, UK \\
}

\pubyear{2019}

\begin{document}
\label{firstpage}
\pagerange{\pageref{firstpage}--\pageref{lastpage}}
\maketitle

\begin{abstract}
  We study the formation of planes of dwarf galaxies around Milky Way
  (MW)-mass haloes in the \eagle{} galaxy formation simulation. We
  focus on satellite systems similar to the one in the MW: spatially
  thin or with a large fraction of members orbiting in the same
  plane. To characterise the latter, we introduce a robust method to
  identify the subsets of satellites that have the most co-planar
  orbits. Out of the 11 MW classical dwarf satellites, 8 have highly
  clustered orbital planes whose poles are contained within a
  $22^\circ$ opening angle centred around
  $(l,b)=(182^\circ,-2^\circ)$. This configuration stands out when
  compared to both isotropic and typical $\Lambda$CDM satellite
  distributions. Purely flattened satellite systems are short-lived
  chance associations and persist for less than $1~\rm{Gyr}$. In
  contrast, satellite subsets that share roughly the same orbital
  plane are longer lived, with half of the MW-like systems being at
  least $4~\rm{Gyrs}$ old. On average, satellite systems were flatter
  in the past, with a minimum in their minor-to-major axes ratio about
  $9~\rm{Gyrs}$ ago, which is the typical infall time of the classical
  satellites. MW-like satellite distributions have on average always
  been flatter than the overall population of satellites in MW-mass
  haloes and, in particular, they correspond to systems with a high
  degree of anisotropic accretion of satellites. We also show that
  torques induced by the aspherical mass distribution of the host halo
  channel some satellite orbits into the host's equatorial plane,
  enhancing the fraction of satellites with co-planar orbits. In fact,
  the orbital poles of co-planar satellites are tightly aligned with
  the minor axis of the host halo. 
\end{abstract}


\section{Introduction}
The Milky Way (MW) satellites have a highly inhomogeneous and
anisotropic phase-space distribution whose origin remains one of
the most baffling cosmological mysteries. All the classical dwarfs and
many of the ultra-faint ones lie on a plane which shows an
unexpectedly high degree of flattening \citep[e.g.][]{Kunkel1976,
 Lynden-Bell1976, Lynden-Bell1982, Kroupa2005,Pawlowski2016}. Many of
the satellites have orbits within this plane
\citep[e.g.][]{Metz2008,Pawlowski2013b,Fritz2018} and the classical
dwarfs have orbits that are more circularly biased than predicted by
the current cosmological model \citep{Cautun2017}. Furthermore, the
plane in which most of the satellites reside is nearly perpendicular
to the MW disc \citep[e.g.][]{Lynden-Bell1982,Libeskind2007,Deason2011,Shao2016}, in
contrast with observations of external galaxies where most satellites
are found within the disc plane of the central galaxy
\citep[e.g][]{Brainerd2005,Yang2006,Agustsson2010,Nierenberg2012}.

Observational studies have reveal that flattened satellite
distributions similar to the MW system are ubiquitous. Our two nearest
giant neighbours, M31 and Centaurus A, both have one or more planes of
satellite galaxies \citep[e.g.][]{Conn2013,Shaya2013,Tully2015}, with
many of their members showing correlated line-of-sight
velocities that potentially a indicate co-rotating configuration 
\citep{Ibata2013,Muller2018,Hodkinson2019}. Farther afield, \citet{Cautun2015b} have
shown that external galaxies also have anisotropic satellite
distributions.

Within the standard cosmological model, the anisotropic distribution
of satellites is a manifestation of the preferential direction of
accretion into haloes
\citep[e.g.][]{Aubert2004,Knebe2004,Zentner2005b,Deason2011,Wang2014,Shi2015,Buck2015}. The
plane of satellites most likely reflects the connection between a
galaxy and its cosmic web. Multiple satellites are accreted along the
same filament \citep{Libeskind2005,Buck2015} which leads to a
significant population of co-rotating satellites
\citep{Libeskind2009,Lovell2011,Cautun2015b}. Correlated satellite
orbits can arise from the accretion of dwarf galaxy groups
\citep[e.g.][]{Li2008,Wang2013,Smith2016,Shao2018}. However, the MW
plane of satellites is unlikely to have originated from the accretion
of most of the satellites either in one group or along one filament
\citep[][see also \citealt{Metz2009,Pawlowski2012b}]{Shao2018}.

Most studies of the MW plane of satellites have focused on the
population of classical satellites because these objects are the ones
with the most precise proper motion measurements
\citep[e.g.][]{Gaia2018} and because we have only a partial census of
fainter satellites, with more than half of the predicted population of
MW ultra-faint dwarfs awaiting discovery \citep[][]{Newton2018}.
While most \lcdm{} haloes have planes of satellite galaxies, the
properties of each plane vary from system to system \citep{Cautun2015}
indicating that the planes encode information about the formation
history of the host. For example, \citet{Shao2016} showed that
while most dark matter haloes are aligned with their central galaxies,
this might not be the case for the MW, and is a consequence of the
Galactic plane of satellites being nearly perpendicular on the MW
disc. A plane of satellites could also indicate a major merger during
the evolution of the host
\citep[e.g.][]{Hammer2013,Smith2016,Banik2018}.

Most \lcdm{} planes of satellite galaxies are transient features, with
their thickness and orientation varying in time
\citep{Bahl2014,Buck2016}. This is a result of many of the members not
moving within the plane and also due to gravitational interactions
between satellites, which have the net effect of diminishing phase
space correlations \citep[e.g.][]{Fernando2017,Fernando2018}. The same
holds true for the Galactic plane of satellites, since several of the
classical dwarfs orbit outside the plane
\citep[e.g.][]{Gaia2018}. Moreover the MW has the Large Magellanic
Cloud (LMC) which is thought to be massive
\citep{Penarrubia2016,Shao2018b,Cautun2019} and thus might also
perturb the obits of the other satellites
\citep[e.g.][]{Gomez2015}. Even in the ideal case, that is when
assuming a spherical MW halo and when neglecting satellite
interactions, the Galactic plane of satellites is short lived and
loses its thinness in ${\sim}1\Gyr$ \citep[][the proper motion errors make the timescale somewhat uncertain -- e.g. see \citealt{Pawlowski2017AN}]{Lipnicky2017}.

In this paper we study the formation and evolution of planes of
satellite galaxies similar to the one observed in our galaxy. For
this, we use the \eagle{} galaxy formation simulation
\citep{Schaye2015} which is ideal for this study since it contains a
large sample of MW-mass haloes with satellite populations similar to
the MW classical dwarfs. We start by identifying analogues of the MW
system in terms of either the thinness or the degree of coherent rotation of
the satellite distribution. We study the stability of the planes of
satellites and how the phase-space distribution of satellites in these
systems compares with that of the overall population of MW-mass
haloes. In particular, we focus the analysis on systems where most
satellites orbit roughly in the same plane, since these are both the
most stable planes and also the ones that contain the largest amount
of information about the accretion history of the MW satellites.

The paper is organised as follows. Section~\ref{sec:simul} reviews the
\eagle{} simulation and describes our sample selection;
Section~\ref{sec:methods} introduces the two methods we use for
identifying planes of satellite galaxies; Section~\ref{sec:result}
presents our results on the formation and evolution of planes of
satellites; we conclude with a short summary in
Section~\ref{sec:conclusions}.

\section{Observational and simulation data} 
\label{sec:simul}
Here we give an overview of the MW data and the galaxy formation
simulation used in our study. We also describe the selection criteria
of our sample of MW-like systems and how we follow the evolution of
these systems across multiple simulation outputs.

\subsection{Observational data}
We study the spatial and orbital distribution of the 11 classical
dwarfs of our galaxy. This choice is motivated by two
considerations. Firstly, it is thought that the classical dwarfs are
bright enough that we have a nearly complete census of
them. Secondly, we need galaxy formation simulations that contain a
large number of MW-mass haloes. Such simulations have limited
resolution, and even state-of-the-art ones, such as \eagle{}, resolve
only the most massive substructures of MW-mass haloes.

We use the sky coordinates, distances and radial velocities of the
classical dwarfs from the \citet{McConnachie2012} compilation. The
satellite proper motion are taken from the \textit{Gaia} DR2 release
\citep{Gaia2018}, except for the Leo I and II satellites, for which we
use the HST proper motions since they have lower uncertainties
\citep{Sohn2013,Piatek2016}. We then transform the satellite
coordinates and velocities to the Galactic Centre reference frame
\citep[for details see][]{Cautun2015}. The thickness of the satellite
plane (see Eq. \ref{eq:tensor}) and the orbital pole directions
are calculated in this Galactic Centre frame.

\subsection{The EAGLE simulation}
We make use of the main cosmological hydrodynamical simulation
(labelled \simRef{}) performed as part of the \eagle{} project
\citep{Schaye2015, Crain2015}. The main \eagle{} run is ideal for this
work since: i) due to its large volume, it contains a large number of
MW-mass haloes, ii) it has a high enough resolution to resolve 
satellites similar to the classical dwarfs and follow their orbits, and iii)
resolves the baryonic processes that affect the orbits of satellites,
such as torquing and tidal stripping due to the presence of a central
galaxy disc \citep[see e.g.][]{Ahmed2017}.

The main \eagle{} run simulates a periodic cube of $100\Mpc{}$ side
length using $1504^3$ dark matter particles and an equal number of
baryonic particles. The dark matter particles have a mass of
$9.7\times 10^6 \Msun$, while the gas particles have an initial mass
of $1.8\times 10^6 \Msun$. \eagle{} assumes a \textit{Planck}
cosmology \citep{Planck2014}
%with cosmological parameters: $\Omega_{\rm m}=0.307, \Omega_{\rm b}=0.04825,\Omega_\Lambda=0.693,h=0.6777,\sigma_8=0.8288$ and $n_{\rm s}=0.9611$. 
and uses galaxy formation models calibrated to reproduce: the stellar mass function, the distribution of galaxy sizes, and the relation between supermassive black hole mass and host galaxy mass.

We make use of the \eagle{} halo and galaxy merger trees described in
\citet{McAlpine2016}. Haloes and galaxies were identified using the
\textsc{subfind} code \citep{Springel2001,Dolag2009} applied to the
full matter distribution (dark matter, gas and stars). It consist of
first identifying friends-of-friends (FOF) haloes using a linking
length of $0.2$ times the mean particle separation \citep{Davis1985},
after which each FOF halo is split into gravitationally bound
substructures. The most massive subhalo is classified as the main halo
and its stellar distribution as the central galaxy. The main haloes
are characterized by the mass, $M_{200}$, and radius, $R_{200}$, that
define an enclosed spherical overdensity of $200$ times the critical
density. The position of each subhalo and galaxy is given by the
particle that has the lowest gravitational potential energy. The
merger tree was built on top of the \textsc{subfind} catalogues using
the \textsc{D-Trees} algorithm \citep{Jiang2014}. The method works by
tracing the most bound particles associated with each subhalo, and
identifying in the subsequent simulation outputs the subhalo which
contains the largest fraction of these particles.

\subsection{Sample selection}
To identify systems similar to the MW, we start by selecting the 3209
present day haloes with mass,
$M_{200} \in [0.3, 3] \times 10^{12}\Msun$. The wide mass range is
motivated by the large uncertainties in the mass of the MW
\citep[e.g. see Fig. 7 of][]{Callingham2018} and the need to have a
large sample of such systems. We require that any such halo be
isolated and thus we remove any galaxy that has a neighbour within
$600\kpc$ that has a stellar mass larger than half their mass. We also
restrict our selection to systems that, like the MW, have at least 11
luminous satellites within a distance of $300 \kpc$ from the central galaxy. 
We define luminous satellites as any subhaloes that has at least one stellar
particle associated to them; this corresponds to objects with stellar mass
higher than ${\sim}1\times10^6 \Msun$. We find 1080 host haloes that satisfy
all three selection criteria, with the resulting sample having a median halo
mass, $M_{200} \sim 1.2 \times 10^{12}\Msun$ (the distribution of
host halo masses is shown in Fig. A1 of \citealt{Shao2016}).

To study the evolution of satellite systems, we make use of the
\eagle{} snipshots, which are finely spaced (about every 70 Myrs)
simulation outputs that allow us to trace the orbits of satellites
with very good time resolution. For each satellite and its central
galaxy, we trace their formation history using the most massive
progenitor in the merger trees. Starting at high redshift, we follow
forward the merger tree of each satellite in tandem with the merger
tree of its present day central galaxy, until we find the first
snapshot where the satellite and the central are part of the same FOF
group; we then define the epoch of that snapshot as the infall time
for the satellite. In a small number of cases satellite galaxies may
drift in and out of the host FOF halo. Even in those cases, we define
the accretion time as the first time the satellite enters the
progenitor of the $z=0$ host halo.

\section{Methods}
\label{sec:methods}
Here we describe the two approaches we use to identify analogues of 
the MW planes of satellite galaxies: i) using the spatial distribution
of satellites, which leads to determining \MWthin{} planes, and ii)
using the orbital pole distribution, which leads to determining
\MWorbit{} planes.

\subsection{\textit{MW-like-thin} planes of satellite galaxies}
\label{sec:thin_planes}

We wish to identify planes of satellite galaxies that have a similar
spatial distribution to the Galactic classical dwarfs, which we refer
to as \MWthin{} planes. To find Galactic analogues, we calculate the
thickness of the satellite systems using the mass tensor,
\begin{equation}
    I_{ij} \equiv \sum_{k=1}^{N} x_{k,i}\;x_{k,j}
    \label{eq:tensor} \;,
\end{equation}
where the sum is over the $N=11$ most massive satellites by stellar
mass (hereafter the ``top 11''). The quantity $x_{k,i}$ denotes the
$i$-th component ($i=1,2,3$) of the position vector of satellite $k$
with respect to the central galaxy. The shape and the orientation are
determined by the eigenvalues, $\lambda_i$
($\lambda_{1}\geqslant\lambda_{2}\geqslant\lambda_{3}$), and the
eigenvectors, $\uvec{e}_i$, of the mass tensor. The major,
intermediate and minor axes of the corresponding ellipsoid are given
by $a=\sqrt{\lambda_1}$, $b=\sqrt{\lambda_2}$, and
$c=\sqrt{\lambda_3}$, respectively. We refer to $c/a$ as the thickness
of the satellite system and to $\uvec{e}_3$, which points along the
minor axis, as the normal to the plane of satellites.

%%%%%%%%%%%%%%%%%%%%%%%%%%%%%%%%%%%%%%%%%%%%%%%%%%%%%%%%%%%%%%%%%%%%%
\begin{figure}
    \centering
	\plotone{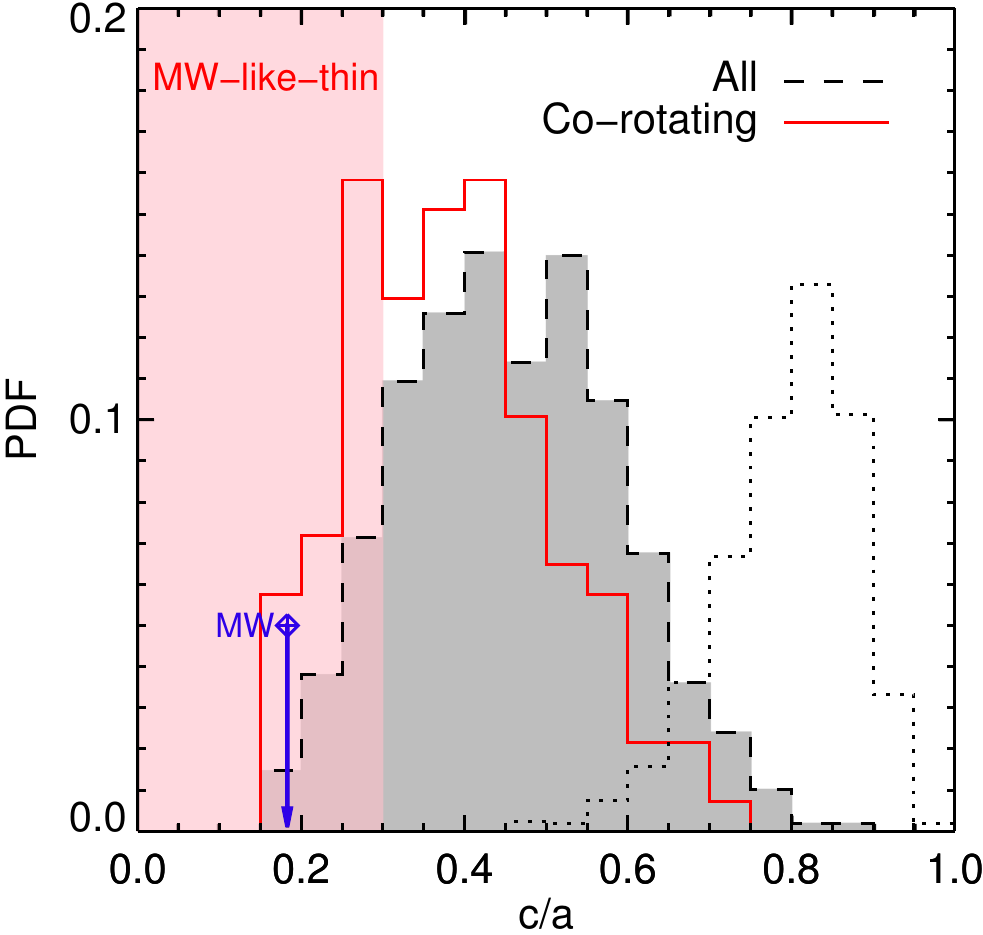}
	\caption{ The axis ratio, $c/a$, distribution for the 11 most
          massive satellites of \eagle{} MW-mass host haloes at
          $z=0$. The dashed line corresponds to all haloes, while the
          solid line shows the hosts with an abundance of co-rotating
          satellites (see Sec. \ref{sec:orbit_planes}), that is hosts
          for which at least 8 of the 11 satellites have orbital
          planes within a $35^\circ$ opening angle. The vertical
          arrow shows the Galactic value, $c/a=0.183$. The red shaded
          region indicates systems with $c/a<0.3$, which represent our
          \eagle{} sample of \MWthin{} systems. The grey dotted line
          shows the $c/a$ distribution of the mass for MW-mass dark
          matter haloes (to better fit the plot, the halo $c/a$ PDF is
          normalized to 0.5 and not to unity). }
	\label{fig:Pdf_ca}
	\vspace{-0.2cm}
\end{figure}
%%%%%%%%%%%%%%%%%%%%%%%%%%%%%%%%%%%%%%%%%%%%%%%%%%%%%%%%%%%%%%%%%%%%%

The distribution of plane thicknesses, $c/a$, for the top 11 satellites
of MW-mass hosts is shown in \reffig{fig:Pdf_ca}. The satellite systems
have a large spread in $c/a$ values ranging from ${\sim}0.15$ to ${\sim}0.9$
and a most likely value of $c/a \approx 0.45$. For comparison, we also
show the shape, $c/a$, of their haloes, which is calculated by applying
\refeq{eq:tensor} to the distribution of dark matter particles within
$R_{\rm 200}$ from the halo centre (see dotted line in \reffig{fig:Pdf_ca}).
On average, the satellites are more flattened and have a wider distribution of $c/a$ values
than their host haloes \citep[e.g. see also][]{Libeskind2005,Kang2005}. 

The 11 classical dwarfs of the MW have an axis ratio, $c/a=0.183$,
which is shown by the vertical arrow in \reffig{fig:Pdf_ca}. This
value is very low when compared with the typical expectation in
\eagle{}, with only ${\sim}1$ percent of \eagle{} MW-mass haloes
having thinner satellite distributions \citep[see
also][]{Wang2013,Pawlowski2014c}. To obtain analogues to the MW planes
of satellites, we select \eagle{} systems with $c/a<0.3$, which
represents our sample of \MWthin{} planes. There are 134 such systems
and they represent $12$ percent of the total sample of \eagle{}
MW-mass haloes. Note that while most of the \MWthin{} planes are
thicker than the MW one, the thickness of the MW plane of satellites
is predicted to increase rapidly with time
\citep[e.g. see][]{Lipnicky2017} and thus our selection procedure is
reasonable.

\subsection{\textit{MW-like-orbit} planes of satellite galaxies}
\label{sec:orbit_planes}

%%%%%%%%%%%%%%%%%%%%%%%%%%%%%%%%%%%%%%%%%%%%%%%%%%%%%%%%%%%%%%%%%%%%%
\begin{figure*}
	\vspace{-0.4cm}
    \plotx{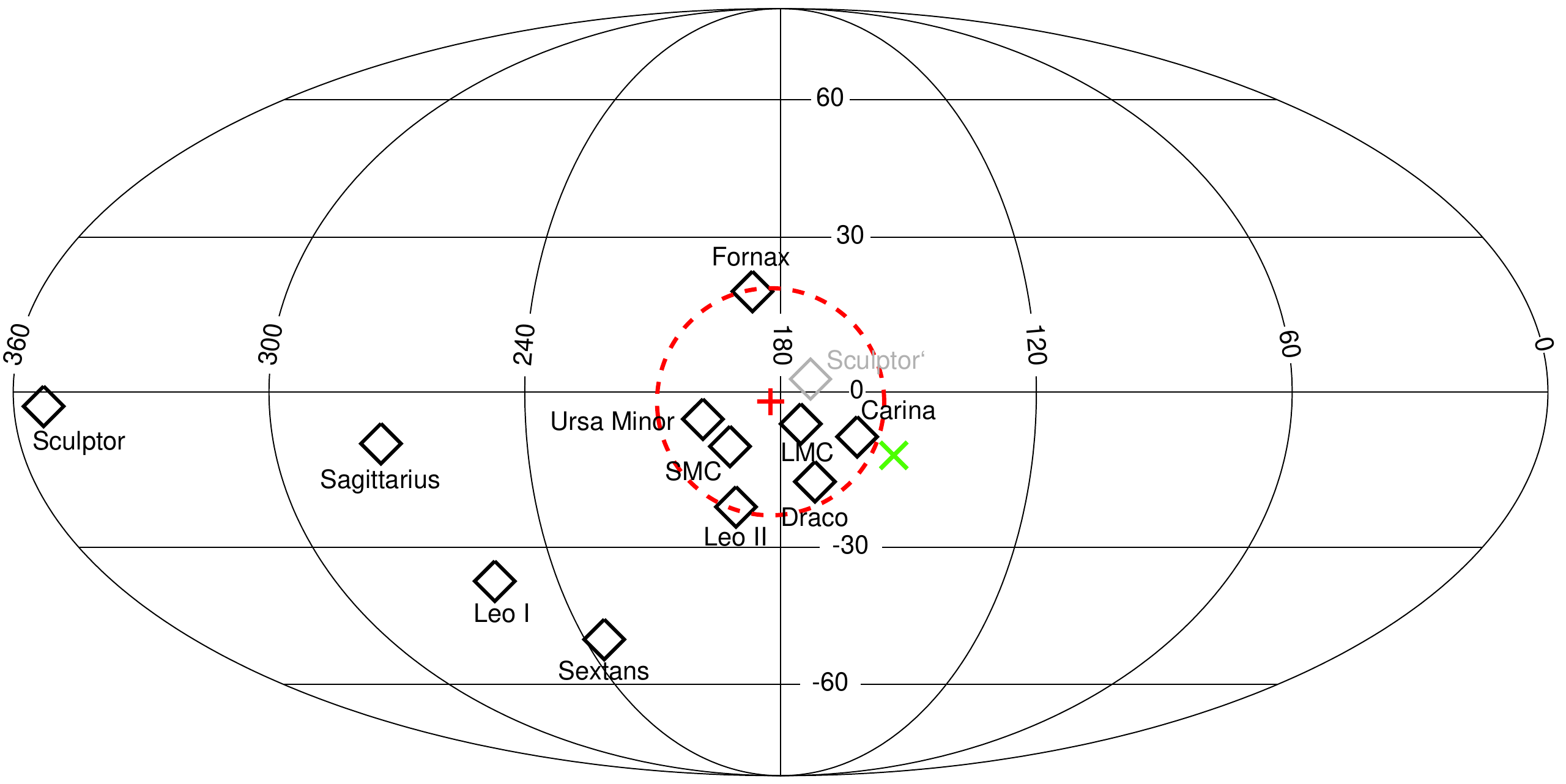}
    \caption{ Aitoff projection of the orbital poles of the MW
      classical satellites. Each black rhombus corresponds to the
      orbital pole of a satellite in Galactic longitude, $l$, and
      latitude, $b$. Out of the 11 classical satellites, 8 orbit in
      roughly the same plane, that is within a $22^\circ$ opening
      angle. The normal to this orbital plane is indicated by the red
      cross symbol and the dashed line shows a $22^\circ$ opening
      angle around this direction. Out of the 8 satellites with
      co-planar orbits, Sculptor is counter-rotating and, to emphasize
      its membership in the orbital plane, the grey rhombus shows its
      position after flipping its orbital pole. The green x symbol
      shows the orientation of the minor axis of the spatial
      distribution of satellites.}
	\vspace{-0.1cm}
	\label{fig:MW_sky}
\end{figure*}
%%%%%%%%%%%%%%%%%%%%%%%%%%%%%%%%%%%%%%%%%%%%%%%%%%%%%%%%%%%%%%%%%%%%%
%%%%%%%%%%%%%%%%%%%%%%%%%%%%%%%%%%%%%%%%%%%%%%%%%%%%%%%%%%%%%%%%%%%%%
\begin{figure}
	\plotone{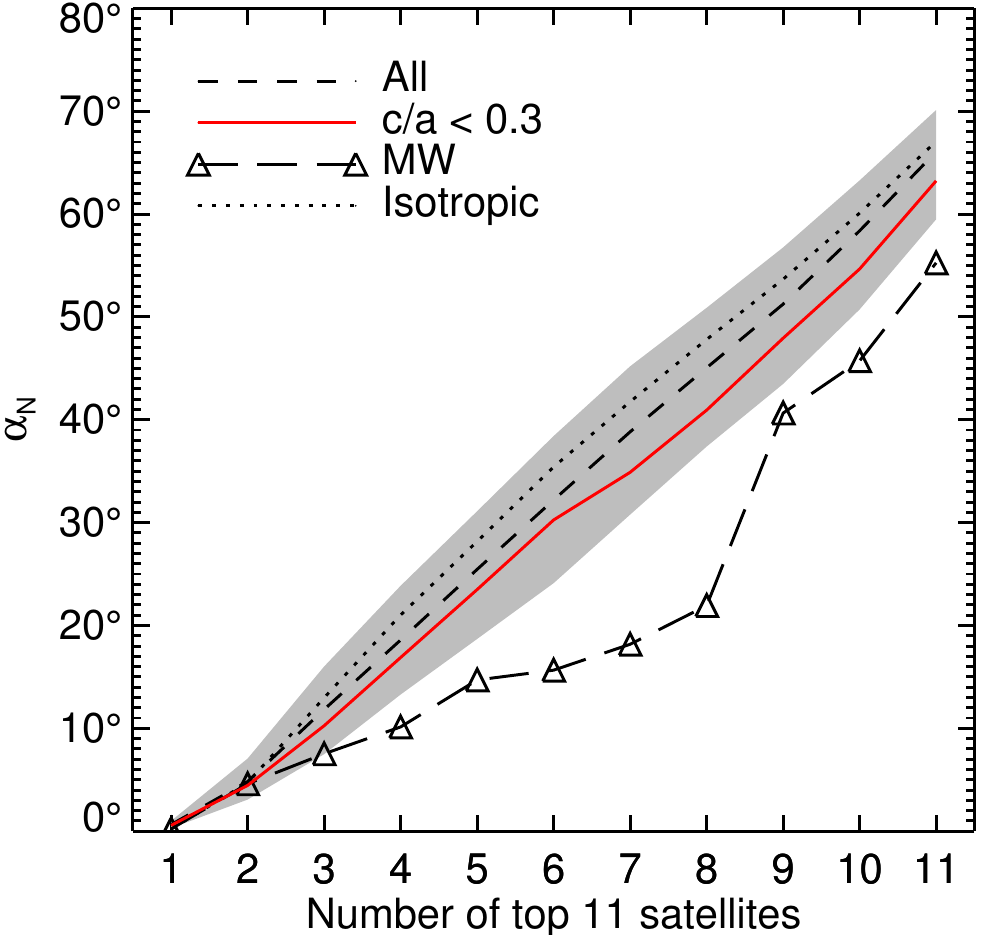}
	\caption{ The distribution of opening angles, $\alpha_N$,
          corresponding to the $N$ out of 11 satellites which have the
          most co-planar orbits. The dashed line shows the median value
          for all MW-mass systems in \eagle{} and the grey shaded
          region shows the 16 to 84 percentiles. The dashed line with
          symbols shows the Galactic classical satellites, with the MW
          having 8 objects with very co-planar orbits. The solid line
          shows the median expectation for \MWthin{} systems,
          i.e. with $c/a<0.3$. The dotted line shows the median
          expectation for isotropically distributed orbital planes.}
	\vspace{-0.2cm}
	\label{fig:Alpha_n}
\end{figure}
%%%%%%%%%%%%%%%%%%%%%%%%%%%%%%%%%%%%%%%%%%%%%%%%%%%%%%%%%%%%%%%%%%%%%
%%%%%%%%%%%%%%%%%%%%%%%%%%%%%%%%%%%%%%%%%%%%%%%%%%%%%%%%%%%%%%%%%%%%%
\begin{figure}
	\plotone{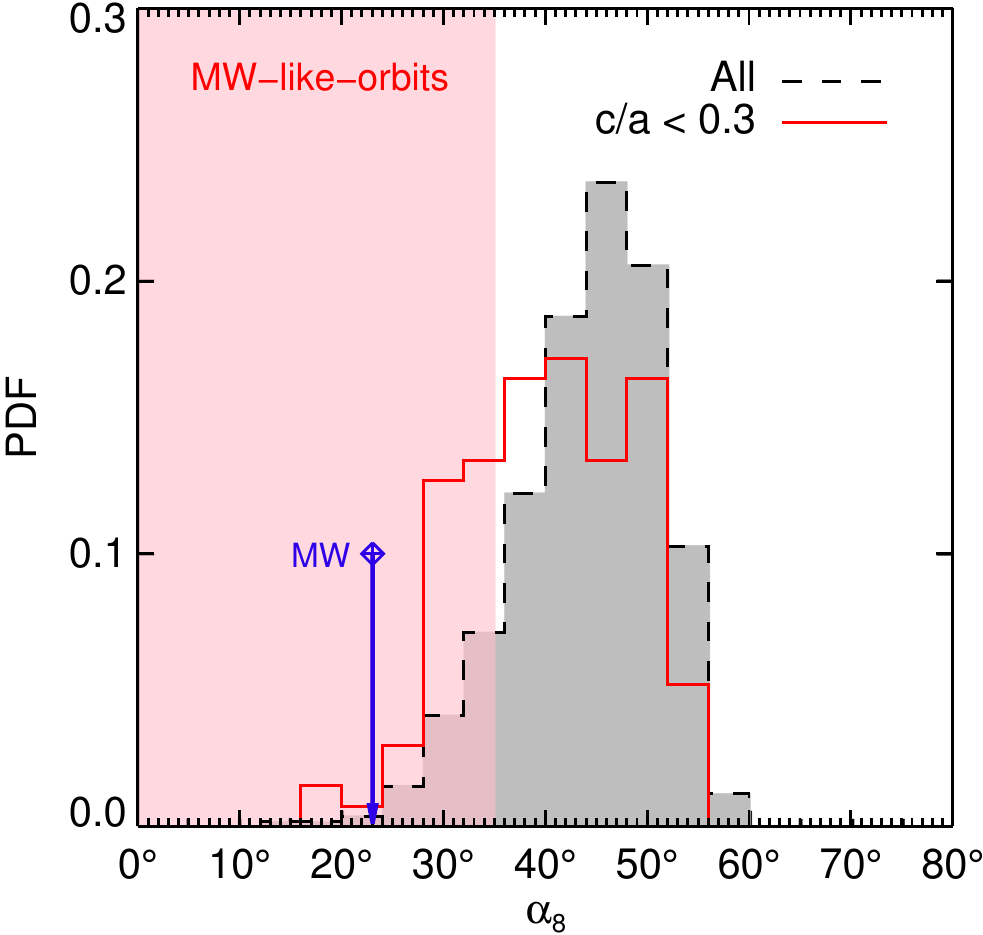}
	\caption{ The distribution of opening angles, $\alpha_8$,
          corresponding to 8 out of the top 11 satellites with the
          most co-planar orbits. The dashed and solid lines correspond
          to all and $c/a{<}0.3$ MW-mass systems in \eagle{}. The
          vertical line indicates the Galactic value,
          $\alpha_8=22^\circ$. The red shaded region corresponds to
          the selection of \MWorbit{} systems, that is those with
          $\alpha_8{<}35^\circ$.}
	\vspace{-0.3cm}
	\label{fig:Pdf_a8}
\end{figure}
%%%%%%%%%%%%%%%%%%%%%%%%%%%%%%%%%%%%%%%%%%%%%%%%%%%%%%%%%%%%%%%%%%%%%

%%%%%%%%%%%%%%%%%%%%%%%%%%%%%%%%%%%%%%%%%%%%%%%%%%%%%%%%%%%%%%%%%%%%%
\begin{table}
  \caption{ The distribution of the orbital poles of the classical satellites, given in terms of Galactic longitude, $l$, and latitude, $b$. The first 8 satellites in this table orbit in approximately the same plane whose normal is given by $(l,b)=(182.3^\circ,-1.8^\circ)$ (see \reffig{fig:MW_sky}). The last column gives the angle, $\Delta$, between each orbital pole and the normal to this plane (see Eq. \ref{eq:direction_angle}). The satellites are ordered in ascending order of the $\Delta$ values. 
  }
    \label{tab:PP}
    \begin{center}
    \begin{tabular*}{0.8\linewidth}{@{\extracolsep{\fill}}lrrl}
    \hline
    \hline
    Name & $l [^\circ]$ & $b [^\circ]$ & $\Delta [^\circ]$ \\
    \hline \\[-.2cm]
        \multicolumn{4}{c}{\bf Satellites orbiting in nearly the same plane} \\
        \hline \\[-.3cm]
        % PP          &  182.3 & -1.8  & 0     \\
LMC         &  175.3 & -6.0  & 8.2   \\
Sculptor$^\dagger$    &  353.0 & -2.7  & 10.3  \\
SMC         &  192.0 & -10.3 & 12.9  \\
Ursa Minor  &  198.3 & -5.1  & 16.3 \\
Draco       &  171.7 & -17.1 & 18.5 \\
Carina      &  161.9 & -8.5  & 21.3 \\
Fornax      &  186.8 &  19.3 & 21.6 \\
Leo II      &  191.0 & -22.0 & 21.9 \\

    \hline \\[-.2cm]
        \multicolumn{4}{c}{\bf Satellites orbiting outside the plane} \\
        \hline \\[-.3cm]
        Sextans     &  234.1 & -49.4 & 64.8 \\
Leo I       &  257.0 & -36.9 & 76.8 \\
Sagittarius$^\dagger$ &  274.6 & -9.8  & 88.0 \\

    \hline
    \end{tabular*}
    \end{center}
% \small \smallskip
$^\dagger$Sculptor and Sagittarius have counter-rotating orbits with respect to the plane normal. The other satellites have co-rotating orbits.
\end{table}
%%%%%%%%%%%%%%%%%%%%%%%%%%%%%%%%%%%%%%%%%%%%%%%%%%%%%%%%%%%%%%%%%%%%%

The MW classical satellites show a surprisingly high degree of
coherent rotation, with many dwarfs orbiting in nearly the same plane
(see \reffig{fig:MW_sky}). Here, we want to better characterise the
orbital structure of the Galactic satellites and to identify similar
satellite configurations in cosmological simulations. To do so, we
start from the question: how many MW satellites orbit approximately in
the same plane and which ones are those?

To answer the question, we first identify which subset of $N$ out of
11 classical satellites has the most planar orbits, for $N$ varying
from 1 to 11. For each value of $N$, we calculate the direction which
contains within the smallest opening angle the orbital poles of $N$ of
the 11 satellites. We find this preferred direction by generating
$10^4$ uniformly distributed points on the unit sphere, which
correspond to $10^4$ uniformly distributed directions. For each
direction, $\uvec{n}_i$, we calculate its angle,
\begin{equation}
    \Delta_{ij} = \arccos\left(~| \uvec{n}_i \cdot \uvec{L}_j |~\right)
    \label{eq:direction_angle} \;,
\end{equation}
with the direction of the orbital angular momentum, $\uvec{L}_j$, of
each of the 11 satellites. We take the absolute value of the
$\uvec{n}_i \cdot \uvec{L}_j$ scalar product to account for the fact
that satellites can be both co- and counter-rotating in the same
plane. The $N$ of the 11 satellites that come closest to rotating in
the plane perpendicular to $\uvec{n}_i$ are the $N$ galaxies with the
smallest $\Delta_{ij}$ angle. The largest of those $N$ angles
determines the minimum opening angle, $\alpha_{N,i}$, around the
$\uvec{n}_i$ direction needed to enclose the orbital poles of all
those $N$ satellites. In practice, we implement this procedure by
sorting the satellites in ascending order of their $\Delta_{ij}$
angle, with the sequence of angles corresponding to the minimum
opening angle, $\alpha_{N,i}$ (with $N$ from 1 to 11). We repeat this
procedure for each of the $10^4$ uniformly distributed directions to
obtain a set of values $\{\alpha_{N,i}\}$ with
$N=1 \rm{~to~} 11$ and $i=1 \rm{~to~} 10^4$.

The subsample of $N$ satellites which have the most co-planar orbits
is obtained by finding the minimum over $i$ of the $\alpha_{N,i}$
values. We define the corresponding direction as the direction of the
plane in which those $N$ satellites rotate. The minimum opening angle,
$\alpha_N \equiv \min_{i} \alpha_{N,i}$, describes how planar are the
orbits of those satellites. The smaller $\alpha_N$ is, the more
clustered are the orbital poles of the $N$ satellites.

To determine the optimal value of $N$ for the MW, we study in
\reffig{fig:Alpha_n} the dependence of the $\alpha_N$ opening angle on
$N$. This is shown in the figure by the dashed line with symbols. The
$\alpha_N$ variation with $N$ for the MW system shows a curious trend:
it increases slowly for small $N$ and then exhibits a large rise as
$N$ is increased from 8 to 9. This is indicative of the MW having 8
classical satellites with roughly co-planar orbits, while the
remaining 3 classical satellites orbit outside this plane. The same
conclusion can be reached when comparing the Galactic orbits with the
median expectation for an isotropic distribution of orbits (see dotted
line in \reffig{fig:Alpha_n}). For $N\leq8$, the MW $\alpha_N$ angle
grows more slowly with $N$ than the median expectation for isotropic
orbits, while for $N>8$, the slope of the two functions is roughly the
same.

\reffig{fig:Alpha_n} also shows the distribution of $\alpha_N$ values
for the brightest 11 satellites of MW-mass haloes in \eagle{}. This
was obtained by applying to each \eagle{} satellite system the same
procedure as for the MW satellites. The figure shows the
median value of $\alpha_N$ (dashed line) and the 16 to 84 percentiles
of the distribution (grey shaded region). The median $\alpha_N$ angle
of MW-mass hosts increases gradually with the number of satellites,
$N$, and, at fixed $N$, its value is slightly lower than the median
expectation for an isotropic distribution of orbits, which indicates
that $\Lambda$CDM haloes have an excess of satellites that orbit
roughly in the same plane. The MW $\alpha_N$ values are systematically
below the \eagle{} results, and especially outside the 68 percentile
region, indicating that the MW has a larger degree of co-planar
satellite orbits than typically expected in $\Lambda$CDM. The
difference in $\alpha_N$ opening angles between the \eagle{} sample
and the MW is largest for $N=8$, indicating again that the orbits of 8
of the classical satellites are much more co-planar than expected.

\reffig{fig:MW_sky} presents the distribution of orbital poles for the
11 classical satellites of the MW. The 8 of them with the largest
orbital coplanarity are the ones shown inside the red-dashed circle,
which corresponds to an opening angle $\alpha_8 = 21.9^\circ$. This is
the minimum opening angle need to enclose the orbital poles of those
satellites. The centre of this circle is found at
$(l,b)=(182.3^\circ,-1.8^\circ)$ (see red cross symbol) and
corresponds to the direction of the plane in which most of the MW
satellites orbit currently. The angles between this plane and the
orbital poles of each MW classical satellite are given in
\reftab{tab:PP}. Out of 8 satellites orbiting in the plane, 7 of them
are co-rotating while one, i.e Sculptor, is counter-rotating. To
better emphasise that Sculptor orbits within the plane, the light grey
symbol in \reffig{fig:MW_sky} shows Sculptor's orbital pole if it were
orbiting in the opposite direction. Using pre-Gaia data, the coherent
alignment of 7 to 9 classical satellites has been pointed out by
\citet{Pawlowski2013b}. Note, however, that they employed a different
method of identifying co-orbiting satellites that has one important
difference compared with our approach: it does not allow for
counter-rotating satellites.

Motivated by the analysis of the MW satellites shown in
Figs. \ref{fig:MW_sky} and \ref{fig:Alpha_n}, we define \MWorbit{}
planes as those for which at least 8 out of the 11 satellites orbit in
a narrow plane (i.e. have a small $\alpha_8$ value). We study such
systems in \reffig{fig:Pdf_a8}, which shows the distribution of the
opening angle, $\alpha_8$, for MW-mass haloes in \eagle{}. The
distribution is peaked at $\alpha_8=45^\circ$, and has a slowly
decreasing tail for small $\alpha_8$ values. When comparing to the MW
$\alpha_8$ value (which is show in \reffig{fig:Pdf_a8} by a vertical
arrow), we find that very few \eagle{} systems have satellites with
orbits as planar as in our own galaxy. In fact, less than 1 percent of
\eagle{} systems have $\alpha_8<22^\circ$, which is the Galactic
value. To obtain a reasonable sample of rotating planes of satellites
similar to the MW, we relax the $\alpha_8$ threshold and define
\MWorbit{} planes as those with $\alpha_8<35^\circ$; this corresponds
to roughly $13$ percent of the \eagle{} sample of MW-mass hosts.

\subsection{Spatially thin and orbitally coherent planes of satellites}
The classical satellites of our galaxy are found to have both a
spatially thin configuration and also a majority of members rotating
in nearly the same plane. This raises two intriguing questions: Do
spatially thin satellite distributions have also a high degree of
coherent orbits? and conversely, are satellite systems with many
planar orbits also spatially thin?

We start by considering the first of the two questions. The solid red
line in \reffig{fig:Pdf_a8} shows the distribution of opening angles,
$\alpha_8$, for the \eagle{} \MWthin{} planes, that is those with
$c/a<0.3$. Compared to the full population, \MWthin{} satellite
systems have systematically smaller $\alpha_8$; however, the
difference is small. Only 30 percent of the \MWthin{} planes are
classified as having \MWorbit{} planes (i.e. $\alpha_8 <
35^\circ$). This is only slightly larger than the 13 percent fraction
of the overall population of MW-mass haloes that are classified as
having \MWorbit{} planes. These results indicate that having thin
planes of satellites does not imply also planar orbits. As we will see
in the next section, most spatially thin planes are due to chance
configurations and thus are short lived.

We now study the typical thickness of \MWorbit{} planes of
satellites. The distribution of axes ratio, $c/a$, for systems with
\MWorbit{} planes is shown by the solid red line in
\reffig{fig:Pdf_ca}. Such system are typically thinner than the
overall population, with the PDF shifted towards overall lower $c/a$
values. About 30 percent of the \MWorbit{} planes are found to have
$c/a<0.3$ and thus are classified as \MWthin{} planes. In general,
\MWorbit{} planes are not necessarily thin since to identify them we
required that 8 out of the 11 satellites have nearly planar
orbits. The other 3 satellites can have very different orbital planes
and, at least for part of their orbit, can be found at large distances
from the common orbital plane of the 8 satellites with the most planar
orbits. Such configurations would result in a large minor axis, $c$,
and thus large $c/a$ ratios.

Many of the discussion points of this section apply to the Galactic
plane of classical satellites. At the moment the plane is spatially
thin; however, since at least 3 of the satellites are moving in a
direction nearly perpendicular to the plane, it will stop being so in
the near future (see e.g. the orbital modelling and analysis of
\citealt{Lipnicky2017} and \citealt{Gaia2018}). Thus, when identifying
analogues to the Galactic plane of satellites, it does not make sense
to be overly restrictive by selecting planes that are spatially thin
and that also have a majority of members which orbit in the same
plane. Such a selection would result in a small sample (40 out of the
1080 MW-mass haloes in \eagle{}) without leading to much physical
insight.

\subsection{How common is the MW satellite system?}
The flattening, $c/a$, and the $\alpha_8$ opening angle of the MW
classical satellites is atypical when compared to the \eagle{} simulation.
This can be easily seen from Figures \ref{fig:Pdf_ca} and \ref{fig:Pdf_a8}
which show that our galaxy is in the tail of the distribution. Compared
to the MW, out of our sample of 1080 satellite systems, $9 \; (0.8\%)$
are thinner and $6 \; (0.6\%)$ have smaller $\alpha_8$ values. In particular,
we find only one system that has lower $c/a$ and $\alpha_8$ values than
our galaxy. This raises two crucial questions that we address in this subsection.

Firstly, does the rarity of the MW plane of satellites pose a challenge
to the standard cosmological model? This question has been addressed by
\citet{Cautun2015} who showed that while many $\Lambda$CDM systems have
planes of satellites, no two planes are the same: the number of satellites
in the plane, the plane thickness and the number of members with coherent
rotation vary from system to system. For example, there are many different
ways of obtaining planar satellite orbits that are as infrequent as the
MW case, such as having 9 out of 11 (instead of the MW's 8) satellites with
orbital poles contained within a 26$^\circ$ (compared to the MW $22^\circ$)
opening angle -- this is because the higher number of satellites compensates
for the larger opening angle to result in a similarly uncommon configuration.
This suggests that the MW plane of satellites is just one possible realisation
out of a very diverse population of planes of satellites. Because of this very
diversity, the frequency of a particular configuration of satellites, such as
the MW one, cannot by itself be used to judge the success or failure of a given
cosmological model.

Secondly, to obtain a reasonably large sample of \textit{MW-like} systems
our selection criteria are not excessively fine-tuned to match the exact
MW satellite distribution. In particular, the \MWthin{} sample is defined
adopting $c/a<0.3$ while the MW has $c/a=0.18$, and the \MWorbit{} systems
are defined adopting $\alpha_8<35^\circ$ while for the MW $\alpha_8=22^\circ$.
This raises the question: do our findings apply to the MW given that most
of the systems we study are less extreme? We strongly suspect that is the
case: the same processes that we see in the simulations for these less extreme
systems are likely to have played at least some role in the formation of
the Galactic plane of satellites. However, to answer this question unequivocally
we would need to use hydrodynamic simulations of a much larger volume (${\sim}100$
times larger than \eagle{} to have useful samples) with at least the same
resolution as \eagle{}. We plan to further study this topic once the ongoing
\textsc{eagle-xl} simulation, whose goal is to have a 30 times higher volume
than \eagle{}, is completed.

\section{Results}
\label{sec:result}
In this section we study the time evolution of planes of satellite
galaxies, beginning with a few individual examples and then focusing
on the population of planar structures as a whole.

\subsection{Evolution of individual planes}
\subsubsection{MW-like-thin planes}
\label{sec:thickness}

%%%%%%%%%%%%%%%%%%%%%%%%%%%%%%%%%%%%%%%%%%%%%%%%%%%%%%%%%%%%%%%%%%%%%
\begin{figure}
    \vspace{-0.3cm}
	\plotone{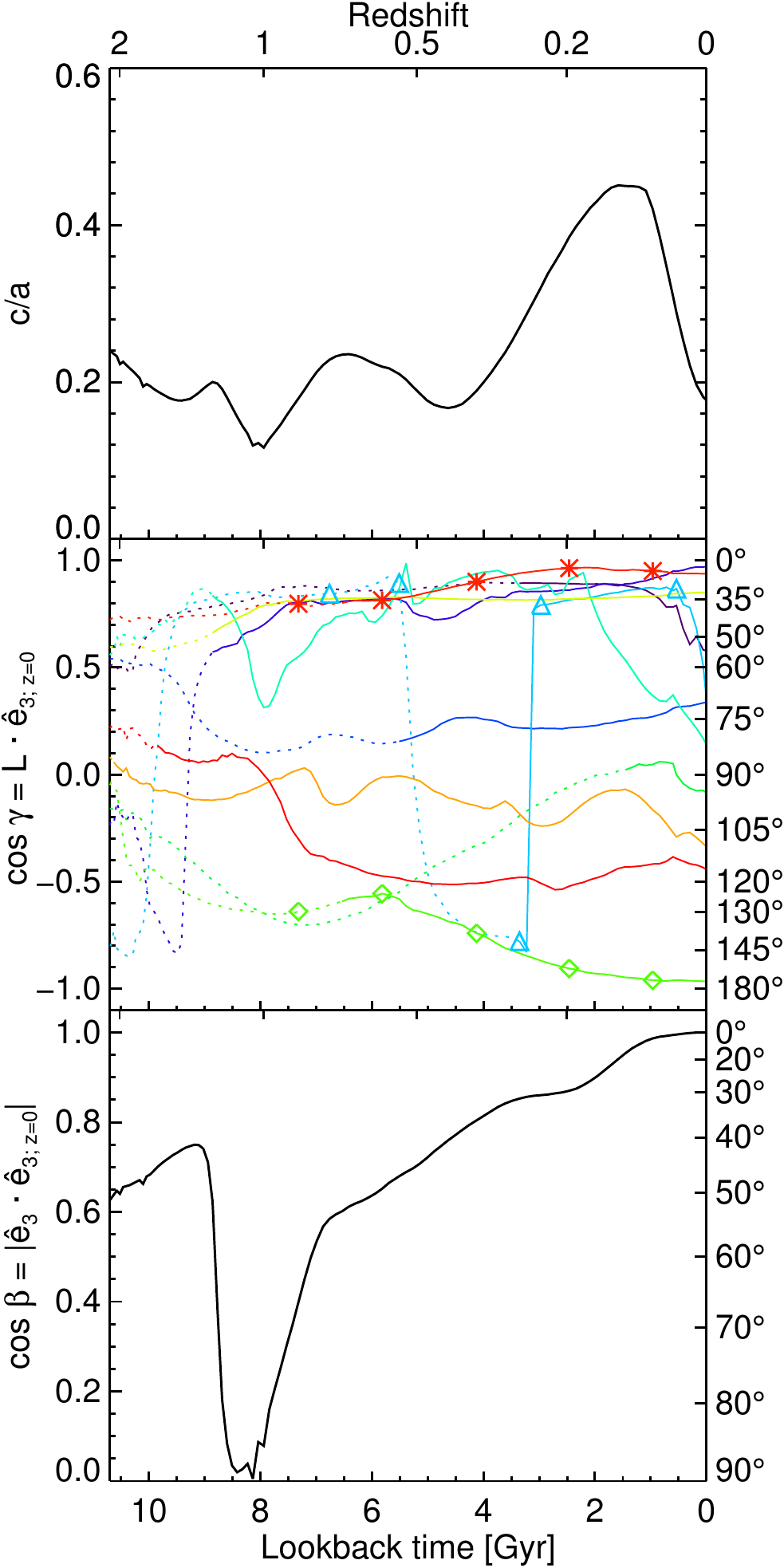}
	\caption{ \textit{Upper panel:} the evolution of $c/a$ for the
          top 11 satellites of a \MWthin{} system from the \eagle{}
          simulation. The satellite system has $c/a=0.176$ at $z=0$,
          similar to the Galactic value of $c/a=0.182$.
          \textit{Middle panel:} the angle between the orientation,
          $\hat{e}_{3,z=0}$, of the $z=0$ plane of satellites and the
          angular momentum, $L$, of each satellite at different
          lookback times. Each colour line shows the main 
          progenitor of each of the top 11 satellites. Each progenitor is
          shown as a dotted line before infall, and as a solid line after
          infall into the MW-mass host halo. The progenitors shown as
          lines with symbols are discussed in detail in the main text.
          \textit{Bottom panel:} the evolution of the alignment angle
          between the orientation of the plane of satellites at $z=0$,
          $\hat{e}_{3,z=0}$, and the orientation at different lookback
          times, $\hat{e}_{3}$.
          }
	\label{fig:MW_ca}
\end{figure}
%%%%%%%%%%%%%%%%%%%%%%%%%%%%%%%%%%%%%%%%%%%%%%%%%%%%%%%%%%%%%%%%%%%%%

%%%%%%%%%%%%%%%%%%%%%%%%%%%%%%%%%%%%%%%%%%%%%%%%%%%%%%%%%%%%%%%%%%%%%
\begin{figure*}
    \vspace{-0.3cm}
	\plottwo{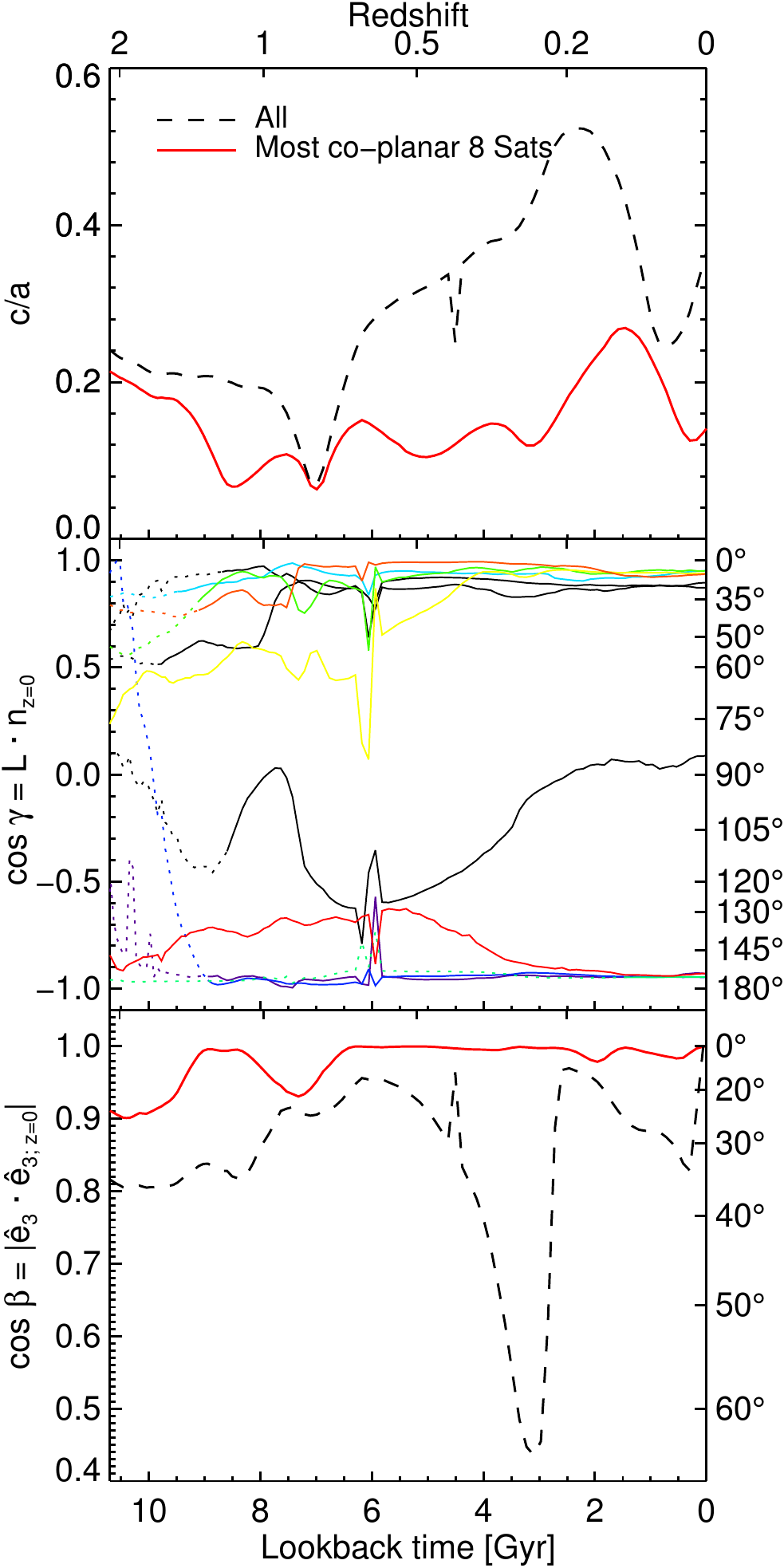}{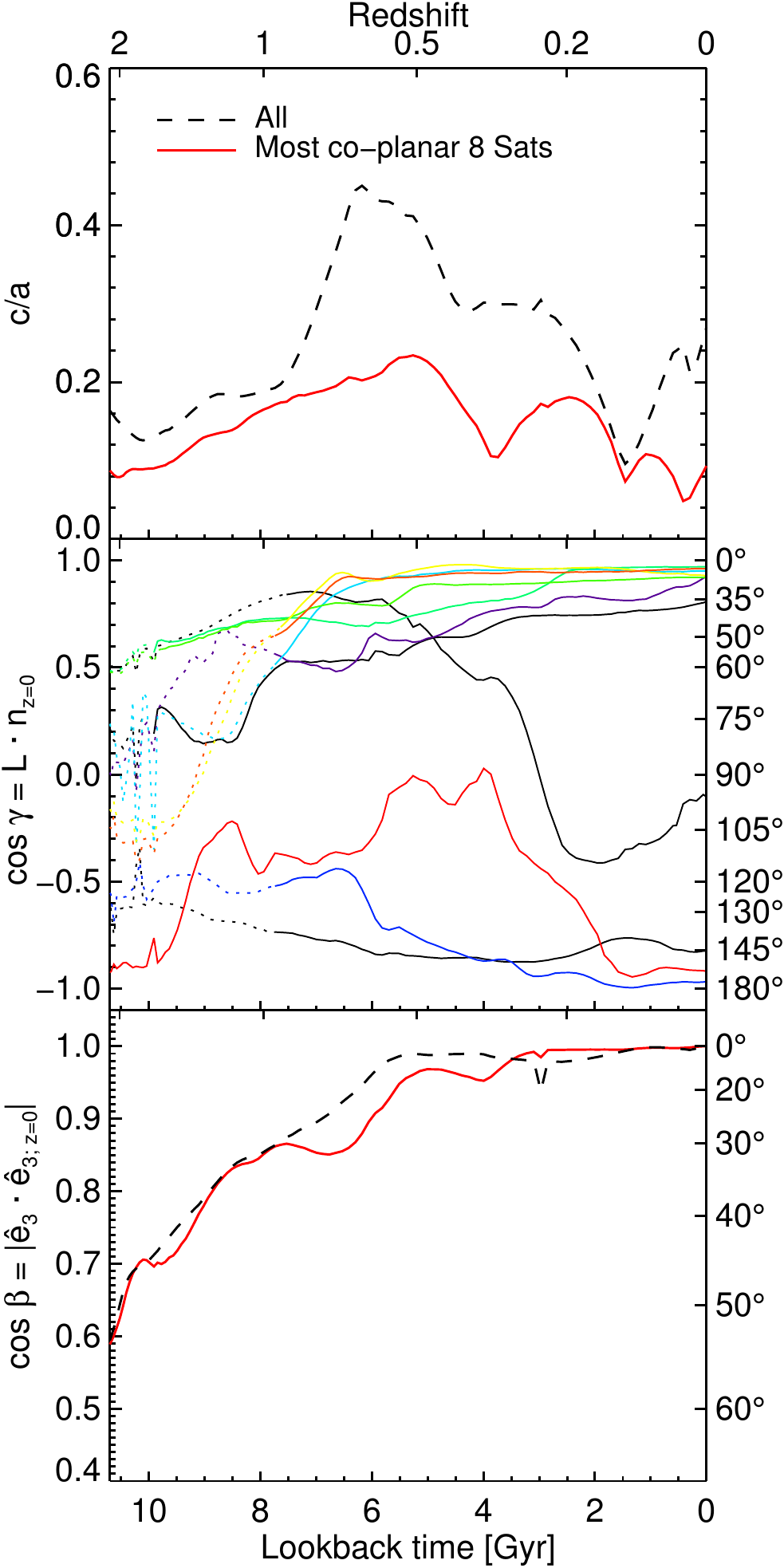}
	\caption{ The evolution of two \eagle{} MW-mass systems which
          have many co-planar satellite orbits at $z=0$. The two
          columns correspond to systems with opening angle,
          $\alpha_8=22^\circ$ (left column) and $\alpha_8=24.5^\circ$
          (right column). \textit{Top panels:} the evolution of the
          axes ratio, $c/a$, for all 11 satellites (black dashed line)
          and for the 8 with the most co-planar orbits (red solid
          line). \textit{Middle panels:} the alignment angle between
          the $z=0$ orientation, $\mathbf{n}_{z=0}$, of the orbital
          plane for the 8 most co-planar satellites and the angular
          momentum, $\mathbf{L}$, of the progenitors of the top 11
          present-day satellites. The colour lines correspond to the 8
          satellites with the most co-planar orbits. \textit{Bottom
          panels:} the alignment angle between the orientation of
          the plane of satellites at $z=0$, $\hat{e}_{3,z=0}$, and the
          orientation at different lookback times, $\hat{e}_{3}$, for
          the top 11 satellites (black dashed line) and for the 8
          satellites with the most co-planar orbits (red solid line).
        }
	\label{fig:examples}
\end{figure*}
%%%%%%%%%%%%%%%%%%%%%%%%%%%%%%%%%%%%%%%%%%%%%%%%%%%%%%%%%%%%%%%%%%%%%

We first study the time evolution of an \eagle{} galactic mass halo
that contains a \MWthin{} plane. We select a satellite system that at
the present time is very flattened, $c/a=0.176$. The flattening is
roughly equal to that of our Galactic satellites, which have
$c/a=0.183$. The evolution of this system is shown in
\reffig{fig:MW_ca}. The top panel shows the axes ratio, $c/a$, for the
progenitors of the top 11 present-day satellites of this system. The
thickness of the system varies rapidly with redshift: $2$ Gyrs ago it
was twice as thick, $c/a>0.4$, while even further ago, it was as thin
as at the present day.

The rapid time evolution of the $c/a$ axes ratio is a consequence of
the complex orbits of the system's members. Many members do not orbit
within the $z=0$ plane of satellites and furthermore the orientation
of their orbital angular momentum can vary in time. This is
highlighted in the middle panel of \reffig{fig:MW_ca}, which shows the
angle, $\gamma$, between the orbital angular momentum of each
satellite and the $z=0$ minor axis, $\hat{e}_{3,z=0}$, of the
satellite system. The evolution of each satellite is shown by a
differently colour line. The line is shown as dotted before infall
and becomes solid when the satellite falls into its host.

We find that some of the satellites have been orbiting in a plane
aligned with the $\hat{e}_{3,z=0}$ vector for at least several
Gyrs. One such example is the top red line with star symbols which
indicates a satellite whose orbital plane has been nearly constant: it
has had a misalignment angle with $\hat{e}_{3,z=0}$ smaller than
$30^\circ$ since at least $10$ Gyrs ago, which is even before its
infall into the host halo $7$ Gyrs ago. However, a significant
fraction of the satellites have an orbital angular momentum whose
direction varies in time. One example is the satellite shown by the
green line with rhombus symbols towards the bottom of the panel: at
$z=0$ it angular momentum vector makes an angle
$\gamma\approx180^\circ$ with $\hat{e}_{3,z=0}$, indicating that it is
counter-rotating in the plane. This satellite was in a nearly
perpendicular orbit $10$ Gyrs ago and, since then, its orbit has been
slowly becoming increasingly aligned with $\hat{e}_{3,z=0}$. An
extreme example of a orbital plane change is shown by the cyan line
with triangle symbols. The angular momentum of this satellite flipped
nearly $180^\circ$ between 5 and 3 Gyrs ago. The flip was due to a
close approach and interaction with a massive satellite (the one in
purple), which changed, at least temporarily, the orbit of the lower
mass satellite.

The example system depicted in \reffig{fig:MW_ca} illustrates that
also the orientation of the plane of satellite orientation can vary
with time. This is highlighted in the bottom panel of \reffig{fig:MW_ca},
which shows the misalignment angle between the
normal, $\uvec{e}_3$, to the satellite distribution at different
epochs and its orientation, $\uvec{e}_{3; z=0}$, at the present day
\footnote{We take the absolute value of the dot product between the
plane of satellite orientations at various times. This is because
the planes are characterised by an orientation and not a
direction. Thus, the misalignment angle can be at most $90^\circ$.}.
While this misalignment angle has been roughly constant during the
last 1~$\rm{Gyr}$, before that it varied rapidly. In particular,
8~$\rm{Gyrs}$ ago the satellite distribution was even more planar than
at the present day (with a minor to major axis ratio, $c/a \sim 0.12$),
however the orientation of that plane of satellites was perpendicular
to the $z=0$ plane of satellites. This indicates that \MWthin{} planes
show a weak coherence between different times.

\subsubsection{MW-like-orbit planes}
\label{sec:po}
We now focus our attention on the evolution of satellites systems that
have the same number of planar orbits as the MW. These are the
\MWorbit{} planes (for details see \refsec{sec:orbit_planes}). We
illustrate the evolution of such planes of satellites by showing in
\reffig{fig:examples} two examples from the \eagle{} simulation that
resemble closely the orbits of the classical MW satellites. The two
examples have 8 out of its top 11 satellites orbiting within an
opening angle, $\alpha_8=22^\circ$ and $24.5^\circ$, respectively.

As was the case of \MWthin{} planes, the flattening of the two
satellite systems varies rapidly with time (see the two top row panels
in \reffig{fig:examples}). While at least 8 of the satellites orbit
within a plane, the other satellites can have very different orbital
planes and the minor-to-major axis ratio, $c/a$, can vary as these
satellites move in and out of the orbital plane shared by the 8 most
co-planar satellites. In fact, the $c/a$ ratio is so sensitive to
individual satellites that even one single object orbiting in a
perpendicular plane can lead to large values of $c/a$. This is the case
for the example shown in the left column of \reffig{fig:examples},
where 2 $\rm{Gyrs}$ ago the satellite distribution had a thickness,
$c/a\sim0.52$, that was predominantly due to one satellite moving in a direction 
perpendicular to the shared orbital plane (middle-row
left-hand panel in \reffig{fig:examples}).

If instead we consider the flattening of the 8 satellites with the
most co-planar orbits, we find an axis ratio, $c/a$, that is
considerably smaller and that varies considerably less with time. This
is illustrated by the solid red line in the two top panels of
\reffig{fig:examples}. Nonetheless, even in this case the $c/a$ ratio
can vary with time since the satellite orbits are not perfectly planar
and, as the satellites approach apocentre, they can find themselves at a larger height from the common orbital plane.

We now focus our analysis on the evolution of orbital planes of
individual satellites, which are shown in the two middle panels of
\reffig{fig:examples}. The lines show the alignment angle between the
orbital angular momentum, $\mathbf{L}$, of a satellite at a given time
and the $z=0$ direction, $\mathbf{n}_{z=0}$, of the common orbital
plane of the 8 satellites with the most co-planar orbits. Alignment
angles of ${\sim}0^\circ$ and ${\sim}180^\circ$ correspond to
satellites that are co-rotating and counter-rotating with the common
orbital plane, respectively. For the system shown in the left-hand
column, most satellites are characterised by very early infall times,
$t<9$ $\rm{Gyrs}$, and have orbital poles that are roughly constant in
time, especially within the last 4 $\rm{Gyrs}$. By contrast, in the
system shown in the right-hand column, the orbital poles of the
satellites were very different at accretion and have converged slowly
towards the present-day distribution when many of them share a common
orbital plane. These two system illustrate the diversity of orbital
histories that can lead to a majority of satellites with co-planar
orbits.

Interestingly, the different orbital evolution of co-planar satellites
translates into distinct predictions for the coherence of \MWorbit{}
planes across cosmic times. This is highlighted in the bottom panels
of \reffig{fig:examples} which show the alignment angle between the
plane of satellites at earlier times and at $z=0$. In particular, the
system shown in the left-hand panel has a roughly constant orientation
for the plane made of the 8 most co-planar satellites. In
contrast, the system shown in the right-hand panel has a plane of
satellites whose orientation has changed over time and only
recently, within the last 6 $\rm{Gyrs}$, has been roughly stable. This
is a manifestation of the present day co-planar satellites having very
different orbital poles at high redshift.

%%%%%%%%%%%%%%%%%%%%%%%%%%%%%%%%%%%%%%%%%%%%%%%%%%%%%%%%%%%%%%%%%%%%%
\begin{figure*}
    \vspace{-0.2cm}
    \plottwotwo{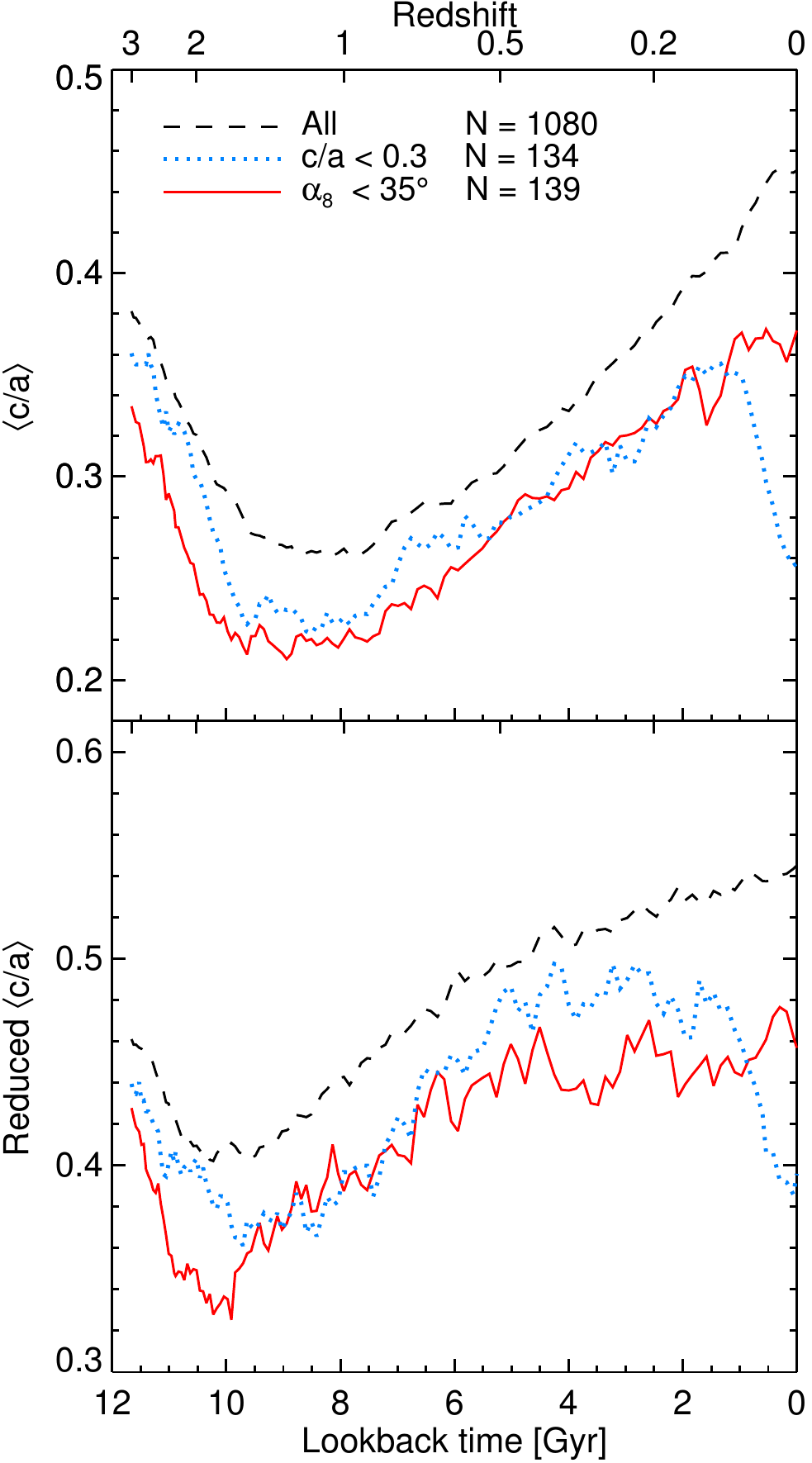}{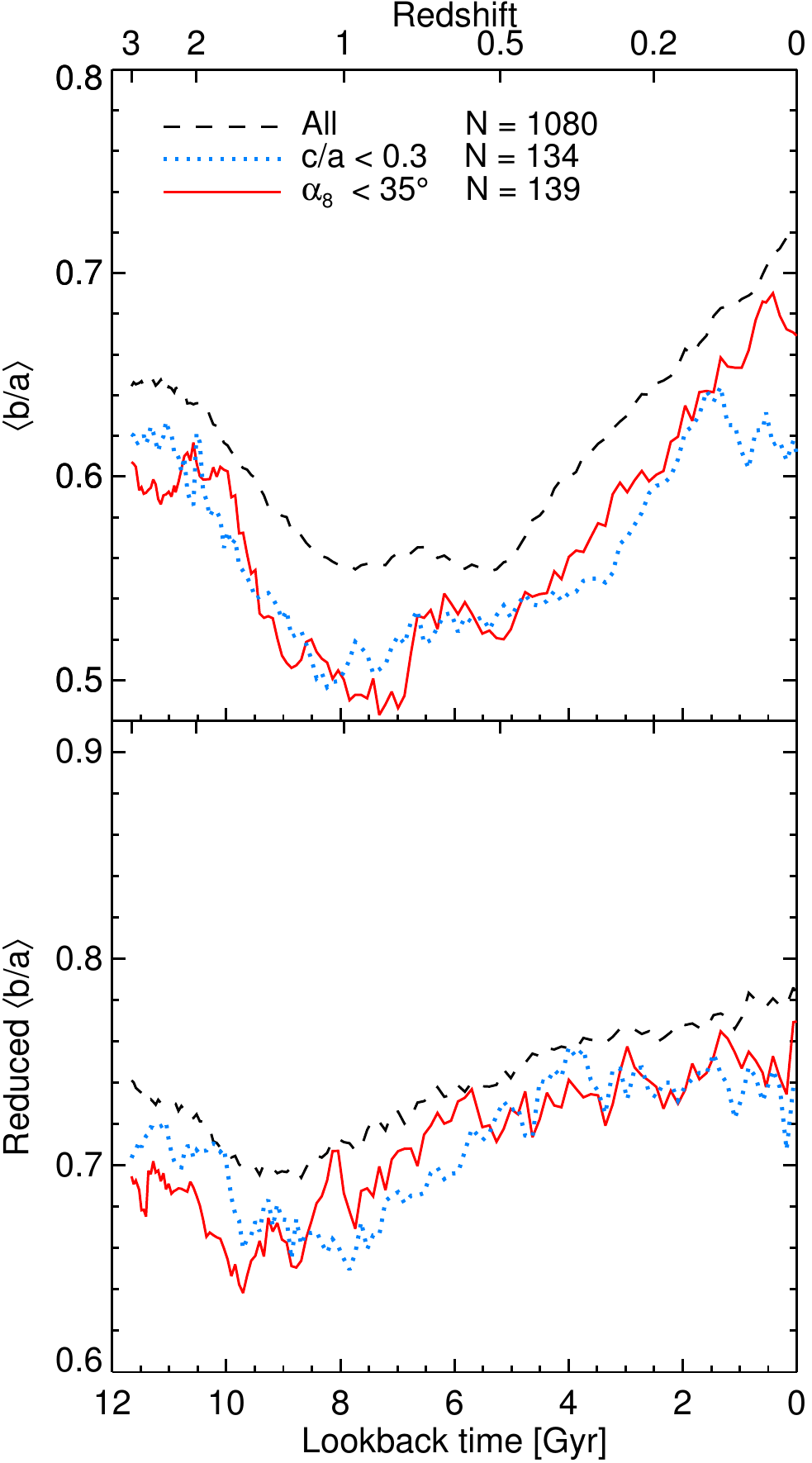}
    \caption{ The time evolution of the median axes ratios,
      $\left\langle c/a \right\rangle$ (left-hand column), and
      $\left\langle b/a \right\rangle$ (right-hand column), for
      satellite systems of MW-mass haloes in the \eagle{}
      simulation. We plot results for the entire population (dashed
      line) as well as for systems identified as having \MWthin{}
      (dotted line) and \MWorbit{} (solid line) planes. The bottom row
      shows $\left\langle c/a \right\rangle$ and
      $\left\langle b/a \right\rangle$ calculated using the reduced
      moment of inertia, that is by scaling each satellite coordinate
      by its distance from the host halo centre. }
	\label{fig:Time_ca}
\end{figure*}
%%%%%%%%%%%%%%%%%%%%%%%%%%%%%%%%%%%%%%%%%%%%%%%%%%%%%%%%%%%%%%%%%%%%%

\subsection{Evolution of the shape and orientation of satellite systems}
In this section we study the formation history of the plane of
satellites population as a whole and assess to what extent the three
satellite systems illustrated in the previous section are typical of
the overall population of planes. As before, we focus the discussion
on \MWthin{} and \MWorbit{} planes, which capture the two defining
characteristics of the Galactic plane of satellites: its thinness and
its high number of satellites with planar orbits.

\subsubsection{The flattening of the overall population of satellite systems}
We start by studying the time evolution of the spatial distribution of
the top 11 brightest satellites. This is illustrated in the top row of
\reffig{fig:Time_ca}, which shows the median values of the axes ratios, $\left\langle c/a \right\rangle$ and $\left\langle b/a \right\rangle$,
as a function of lookback time. The top-left-hand panel shows that the
median $c/a$ for the full sample of MW-mass systems decreases in the
past, and indicates that satellites systems were typically thinner at higher
redshift. The minimum of $\left\langle c/a \right\rangle$ is found at
a lookback time of ${\sim}9~\rm{Gyrs}$ (i.e. redshift $z{\sim}1.3$),
when the satellite distribution was $60\%$ flatter, that is a
$\left\langle c/a \right\rangle=0.27$, compared to the present-day
value of $0.45$. Further in the past, the flattening of the satellite
distribution increases rapidly.

As we have seen in the previous section, the flattening of the
satellite distribution can be affected by just a few satellites. For
example, if the progenitor of one of the present-day satellites is far
from the host (such as before accretion). This can result in a very
large value for the major axis of the system without a corresponding
increase in the value of the minor axis. This would give rise to a
reduction of the axis ratio, $c/a$. To check if the observed trend
with lookback time is due to this effect, we calculate the time evolution
of the ratio between the minor and major axes of the reduced moment of
inertia tensor of the system. This is obtained by modifying
Eq. \eqref{eq:tensor} to include a weight, $w_{i}=1/d_{i}^2$, for each
satellite, where $d_{i}$ is the distance of the satellite from the
central galaxy. This ensures that all satellites contribute equally to
the reduced moment of inertia of the system. The evolution of the
reduced $\left\langle c/a \right\rangle$ axes ratio is shown in the
bottom-left-hand panel of \reffig{fig:Time_ca}. We find the same
qualitative trend, that the median axes ratio decreases towards the past,
with a minimum ${\sim}10~\rm{Gyrs}$ ago. Thus, the minimum seen in the
left-hand column of \reffig{fig:Time_ca} is physical and indicates a
pronounced flattening of the progenitor system at redshift,
$z{\sim}1.5$.

The lookback time corresponding to the minimum of
$\left\langle c/a \right\rangle$ coincides with the typical accretion
time of these satellites; half of the present-day satellites were
accreted more than $8.5~\rm{Gyrs}$ ago \citep{Shao2018}. This paints
an intriguing picture for the evolution of the satellite progenitors:
starting at high redshift, $z>3$, they first move towards the cosmic
web sheet that surrounds the progenitor of their $z=0$ host, and, once
there, they move in the plane of this sheet towards their present-day
host. This is a manifestation of the anisotropic gravitational
collapse of matter, with overdensities first collapsing along one
dimension to form large-scale sheets, then along a second dimension to
form filaments, and finally collapsing along the last direction to
produce virialized haloes
\citep{Zeldovich1970,Arnold1982,Icke1973}. Once inside the host, the
satellite system becomes thicker with time. This thickening could be
due to the satellites moving on different orbital planes already at
the time of accretion, as well as to torques and interactions inside
the host halo that can modify the orbital plane of satellites, as we
have seen in Figs. \ref{fig:MW_ca} and \ref{fig:examples}.
 
The picture of anisotropic gravitational collapse raises an
interesting question: once the satellite progenitors are distributed
in a plane, do they further evolve into a filamentary configurations
before falling into their host halo? We investigate this in the
top-right-hand panel of \reffig{fig:Time_ca}, which shows the time
evolution of the median axes ratio, $\left\langle b/a
\right\rangle$. This is a measure of how filamentary a distribution
is, with small $b/a$ values corresponding to thin filaments. The plot
shows that on average the $b/a$ ratio was smaller in the past, reaching
at minimum ${\sim}7~\rm{Gyrs}$ ago. To interpret this result, we
also need to study the evolution of the same axis ratio but for the
reduced moment of inertia, which is shown in the bottom-right-hand
panel of \reffig{fig:Time_ca}. Interestingly, the
$\left\langle b/a \right\rangle$ ratio for the reduced moment of
inertia shows only a minor decrease at high redshift. This indicates
that the distribution of satellite progenitors does not show strong 
evolution of its filamentary character, that is not all satellites are
accreted along the same filament. This is in agreement with previous
studies which have shown that most MW-mass haloes accrete their
brightest satellites from three or more different filaments
\citep[e.g.][]{Libeskind2005,Gonzalez2016,Shao2018}. Note that,
according to the left-hand column of \reffig{fig:Time_ca}, these
multiple filaments are preferentially found in the same plane
\citep[see also][]{Danovich2012,Libeskind2014,Shao2018}.

\subsubsection{The flattening of systems with planes of satellites}
We now investigate the time evolution of the $c/a$ and $b/a$ axes
ratios for systems that at the present day host a \MWthin{} or
\MWorbit{} plane of satellites. These are shown in \reffig{fig:Time_ca} as dotted and solid
lines, respectively.

By selection, \MWthin{} systems are very thin at $z=0$; however, their
median $c/a$ axes ratio grows rapidly with lookback time. This
indicates that most of these systems are chance alignments of
satellites that do not preserve a low $c/a$ value for an
extended period of time \citep[see
also][]{Bahl2014,Gillet2015,Buck2016}. After the initial increase, the
$\left\langle c/a \right\rangle$ of \MWthin{} systems traces very well
the time variation of $\left\langle c/a \right\rangle$ for the overall
MW-mass population, although the exact value is systematically lower
by ${\sim}0.05$. This indicates that \MWthin{} satellite systems were
on average more planar than the overall population of satellites of
MW-mass haloes at all redshifts. The \MWthin{} sample shows a similar
time evolution in the $\left\langle b/a \right\rangle$ ratio, which,
except for the last $1~\rm{Gyr}$, traces very well the
$\left\langle b/a \right\rangle$ value of the full population, albeit
at a systematically lower value.

\MWorbit{} systems have, at $z=0$, a larger $\langle c/a\rangle$ value than the
\MWthin{} sample; however, since the latter increases rapidly, it
catches up and both have roughly equal $c/a$ values between $1$ and
$6~\rm{Gyrs}$ ago. Even further in the past, the \MWorbit{} systems
are systematically thinner than the \MWthin{} ones. The two samples
have only $30\%$ of their members in common and thus the close match
between the two $c/a$ ratios is unexpected. Even more surprising is
that the $b/a$ axes ratios of the two samples are roughly equal for
lookback times above $1~\rm{Gyr}$.

We also studied the evolution of $\left\langle c/a \right\rangle$ and
$\left\langle b/a \right\rangle$ for the small sample of systems that
fulfil both the \MWthin{} and \MWorbit{} selection criteria. For
clarity, we do not show these results in \reffig{fig:Time_ca}. The
combined systems have the same axes ratio as the \MWthin{} sample for
lookback times less than $1~\rm{Gyr}$, after which their time
evolution matches very closely that of the \MWorbit{} sample, albeit
with some scatter due to the small sample size.

%%%%%%%%%%%%%%%%%%%%%%%%%%%%%%%%%%%%%%%%%%%%%%%%%%%%%%%%%%%%%%%%%%%%%
\begin{figure}
    \vspace{-0.5cm}
	\plotone{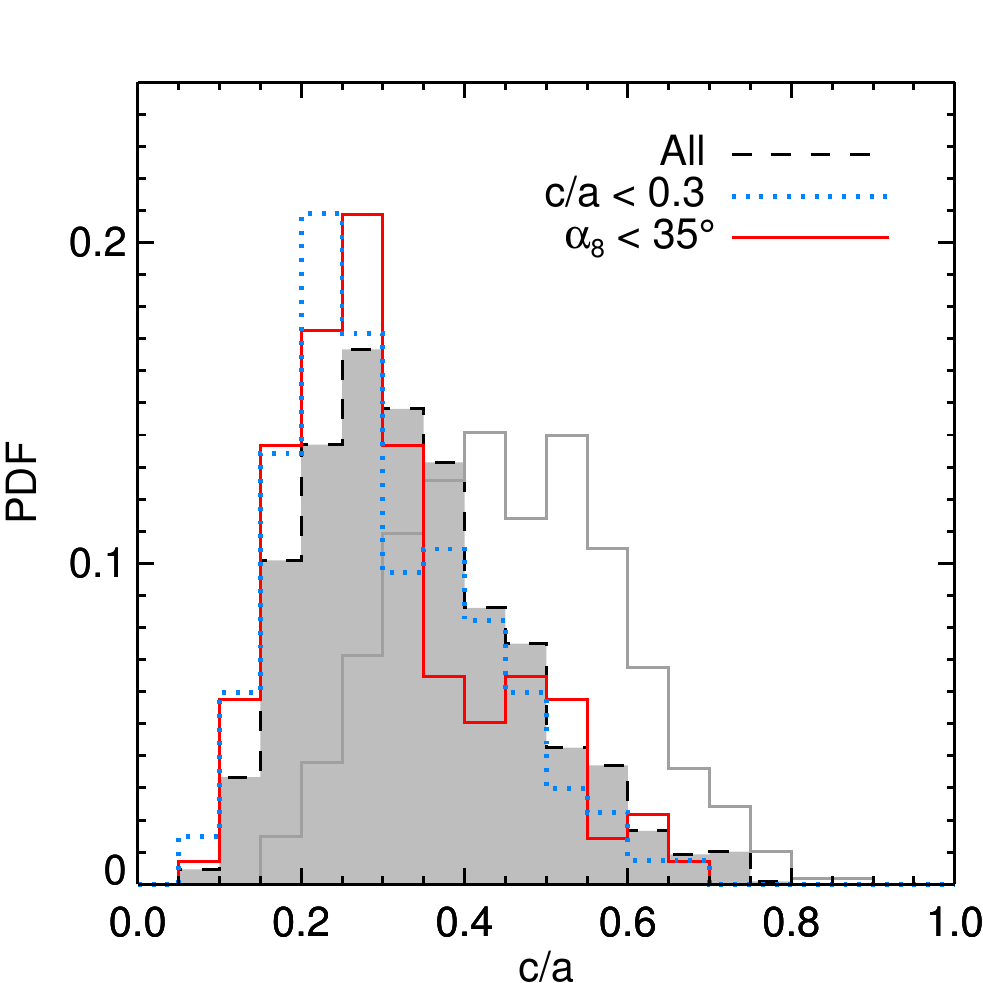}
	\caption{ The distribution of axes ratios, $c/a$, for satellite
          entry points into their $z=0$ MW-mass hosts. The entry
          points are calculated at the time of accretion into the FOF
          halo of the MW-mass host progenitor. The light grey curve
          shows the distribution of the satellite system $c/a$ at
          $z=0$. }
	\label{fig:Time_ca_infall}
\end{figure}
%%%%%%%%%%%%%%%%%%%%%%%%%%%%%%%%%%%%%%%%%%%%%%%%%%%%%%%%%%%%%%%%%%%%%

\subsubsection{The flattening of satellite systems at infall}
Present day planes of satellites correspond to satellite systems that
were systematically thinner at $z>0$ and especially at the time when
most satellites were accreted. This raises an intriguing question: do
planes of satellites form in the host haloes where satellite infall
was most anisotropic? Such a picture has been suggested by previous
studies \citep[e.g.][]{Libeskind2005,Buck2015}. However, since they
employed very small samples of haloes, such studies lacked the
statistical power to give a definitive answer. We investigate this
question in \reffig{fig:Time_ca_infall}, which shows the distribution
of $c/a$ axes ratios for the satellites' entry points into their
host. A satellite's entry point is defined as the position of its
progenitor at the time it first entered the progenitor FOF halo of its
MW-mass host. The entry position is calculated with respect to the
host halo centre at infall time. The distribution of $c/a$ at infall
cannot be directly compared to the results in \reffig{fig:Time_ca}
since the classical satellites of MW-mass hosts have a wide range of
infall times (see Fig. 12 in \citealt{Shao2018}).

\reffig{fig:Time_ca_infall} shows that the galactic satellites formed
a more flattened distribution at infall than at the present time
(compare the dashed with the grey solid lines), and supports our
conclusion that the orbital evolution of satellites typically leads to
more isotropic distributions at $z=0$ \citep[see
also][]{Bowden2013}. In particular, the small $c/a$ values at infall
reflect the anisotropic nature of the large-scale mass distribution
surrounding a MW-mass halo \citep[see
e.g.][]{Lovell2011,Cautun2014c,Libeskind2014,Shao2016}.

The two subsamples with present-day MW-like satellite distributions
had even lower values $c/a$ at infall than the overall population, but
the difference is rather small. In particular, some \MWthin{} and
\MWorbit{} systems had infall $c/a$ ratios of $0.5$ or higher, while
many satellite systems with $c/a\sim0.2$ at infall did not evolve into
present day MW-like satellite populations. This suggests that highly
anisotropic accretion is just one of a number of processes that
lead to the formation of planes of satellite galaxies, and that it
might not even be the dominant factor.

%%%%%%%%%%%%%%%%%%%%%%%%%%%%%%%%%%%%%%%%%%%%%%%%%%%%%%%%%%%%%%%%%%%%%
\begin{figure}
	\plotone{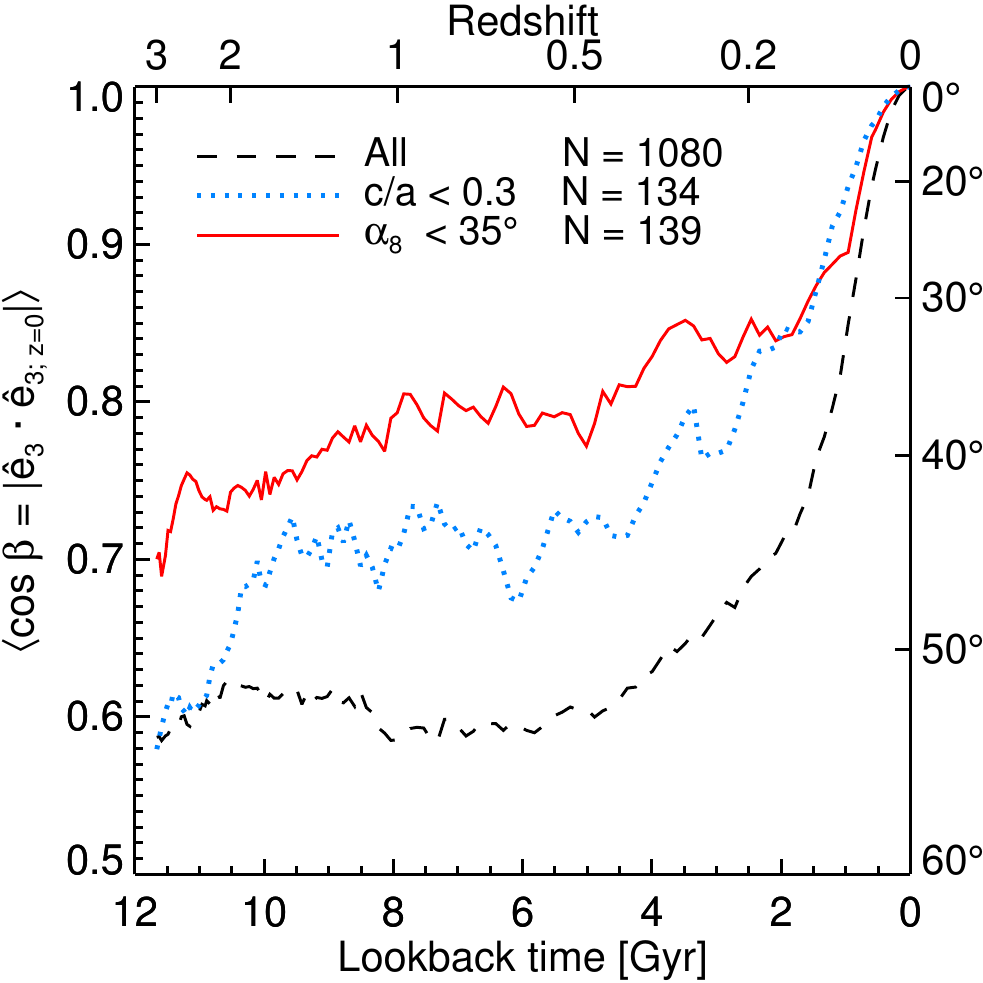}
	\caption{ The directional coherence of the minor axis,
          $\uvec{e}_3$, of the satellite distribution. It shows the
          time evolution of the median misalignment angle, $\beta$,
          between $\uvec{e}_3$ at different lookback times and its
          present-day value, $\uvec{e}_{3;z=0}$. The three lines
          correspond to the overall population of MW-mass haloes as
          well as to $z=0$ systems that have \MWthin{} and \MWorbit{}
          planes of satellites. A complete lack of alignment
          corresponds to $\left\langle \cos\beta \right\rangle=0.5$.
        }
	\label{fig:Time_e3}
\end{figure}
%%%%%%%%%%%%%%%%%%%%%%%%%%%%%%%%%%%%%%%%%%%%%%%%%%%%%%%%%%%%%%%%%%%%%

\subsubsection{The directional coherence of satellite systems}
We now study how the orientation of the satellite systems changes with
time. For this, we consider the eigenvector, $\uvec{e}_3$, that points
along the minor axis of the satellite distribution.
\reffig{fig:Time_e3} shows the evolution of the median misalignment
angle, $\beta$, between $\uvec{e}_3$ at different epochs and its value
at $z=0$. For the full sample, $\beta$ increases rapidly with lookback
time, reaching a value of
$\left\langle \beta \right\rangle {\sim}45^\circ$ at $2~\rm{Gyrs}$,
and then tends to a constant value,
$\left\langle \beta \right\rangle {\sim}55^\circ$, further in the
past. The value of $\left\langle \beta \right\rangle$ is slightly
lower than in the case of no alignment, for which
$\left\langle \beta \right\rangle =60^\circ$, and indicates a weak 
directional coherence between the shape of the present-day satellite
system and that of its progenitors. Thus, in general, both
the axes ratio, $c/a$, and the minor axis orientation, $\uvec{e}_3$,
show a large and rapid variation with time.

The shape of \MWthin{} satellite systems shows more directional
coherence across time than the full population, with the median
misalignment angle at large lookback times tending to a constant
value, $\left\langle \beta \right\rangle {\sim}45^\circ$. The stronger
alignment is due to \MWthin{} systems having a higher fraction of
satellites that share the same orbital plane (see
e.g. \reffig{fig:Pdf_a8}) that helps to preserve the directional
coherence of the distribution. This hypothesis is supported by the
time dependence of $\left\langle \beta \right\rangle$ for the
\MWorbit{} subsample, which shows even better coherence across time
than the \MWthin{} systems. After an initial rapid increase in the
misalignment angle, the \MWorbit{} systems have
$\left\langle \beta \right\rangle {\sim}35^\circ$ that grows slowly
with lookback time.

%%%%%%%%%%%%%%%%%%%%%%%%%%%%%%%%%%%%%%%%%%%%%%%%%%%%%%%%%%%%%%%%%%%%%
\begin{figure}
	\plotone{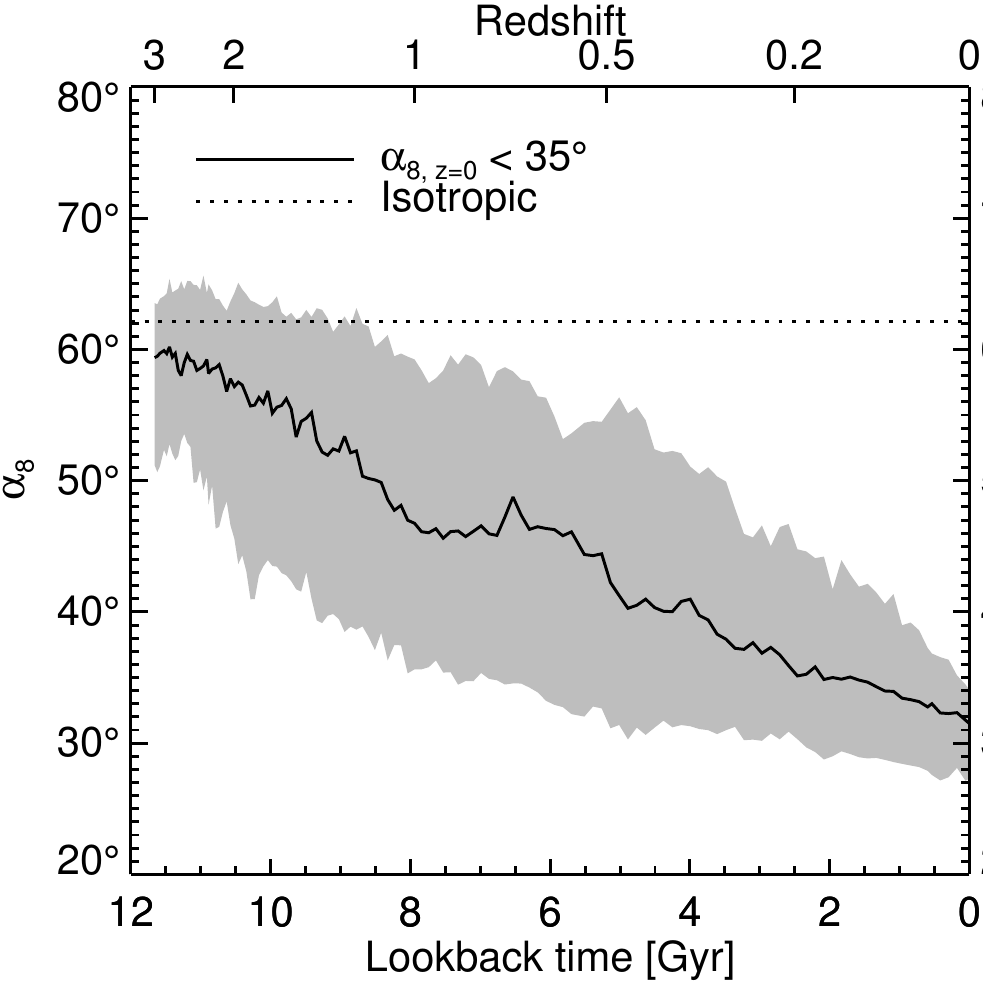}
	\caption{ The time evolution of the opening angle, $\alpha_8$,
          that encloses the orbital poles of 8 of the top 11
          satellites with the most co-planar orbits at the present
          day. We show results only for the subsample with
          $\alpha_8\leq 35^\circ$ at $z=0$. The solid line shows the
          median value of $\alpha_8$ while the shaded region shows the
          16 to 84 percentiles of the distribution. The horizontal
          dotted line marks the mean angle for an isotropic
          distribution of orbits.}
	\label{fig:Time_a8}
\end{figure}
%%%%%%%%%%%%%%%%%%%%%%%%%%%%%%%%%%%%%%%%%%%%%%%%%%%%%%%%%%%%%%%%%%%%%

\subsection{Evolution of the satellites with co-planar orbits}
We now investigate the driving factors that lead to many satellites
having nearly co-planar orbits. In particular, we focus on \MWorbit{}
systems, for which at least 8 of the 11 brightest satellites orbit
within a narrow plane. In \refsec{sec:po} we studied two examples of
such planes: in the first one, the 8 co-planar satellites had been
orbiting in the same plane for the last $8~\rm{Gyrs}$, while in the
second example the co-planar configuration was formed only
recently. In \reffig{fig:Time_a8} we study which of the two examples
describes the typical formation history of a \MWorbit{} system. Here
we plot, as a function of time, the smallest opening angle,
$\alpha_8$, needed to enclose the orbital poles of the 8 satellites
with the most co-planar orbits at the present day.

\reffig{fig:Time_a8} shows that the median value of $\alpha_8$
increases with lookback time, from $31.5^\circ$ at $z=0$ to
${\sim}50^\circ$ at $z=1$. By construction, the median value of
$\alpha_8$ at the present day is low because we are considering only
\MWorbit{} systems which have $\alpha_8\leq35^\circ$ today. We
conclude that, on average, configurations of 8 satellites with
co-planar orbits today must have formed recently. In particular, if we
take the formation time as the point when $\alpha_8$ falls below
$40^\circ$, we find that half of the systems formed less than
$4~\rm{Gyrs}$ ago. However, there is large variability amongst
individual systems, as indicated by the grey shaded region in
\reffig{fig:Time_a8}, which shows the 16 and 84 percentiles of the
distribution. At one extreme, the 16 percentile line remains below
$\alpha_8 = 40^\circ$ for the past $9~\rm{Gyrs}$, while, at the other
extreme, the 84 percentile line is below $\alpha_8 = 40^\circ$ only
for the past $1~\rm{Gyr}$. Thus, there is an important population of
both very old and very young \MWorbit{} planes of satellites. We note
that individual systems show short term variations in $\alpha_8$ on
top of the long-term trend of decreasing $\alpha_8$ with time; this
makes it difficult to determine unambiguously the formation time of a
given plane.

The time variation of the average $\alpha_8$ opening angle is roughly
constant (notwithstanding some variability on very short
timescales). This suggests that the dominant processes which lead to
co-planar satellite orbits are acting consistently over long periods
of time. In \refsec{sec:thickness} we saw that massive satellites can
induce radical changes in the orbital planes of other
satellites. However, while such changes might nudge a satellite orbit
to the common orbital plane of other dwarfs, on average massive
satellites increase the dispersion of orbital poles and thus
contribute to destroying co-planar satellite distributions
\citep[e.g. see][]{Fernando2018}. Moreover, satellite--satellite
interactions are one of the main factors that determine short-term
variability in $\alpha_8$ for individual systems.

%%%%%%%%%%%%%%%%%%%%%%%%%%%%%%%%%%%%%%%%%%%%%%%%%%%%%%%%%%%%%%%%%%%%%
\begin{figure}
	\plotone{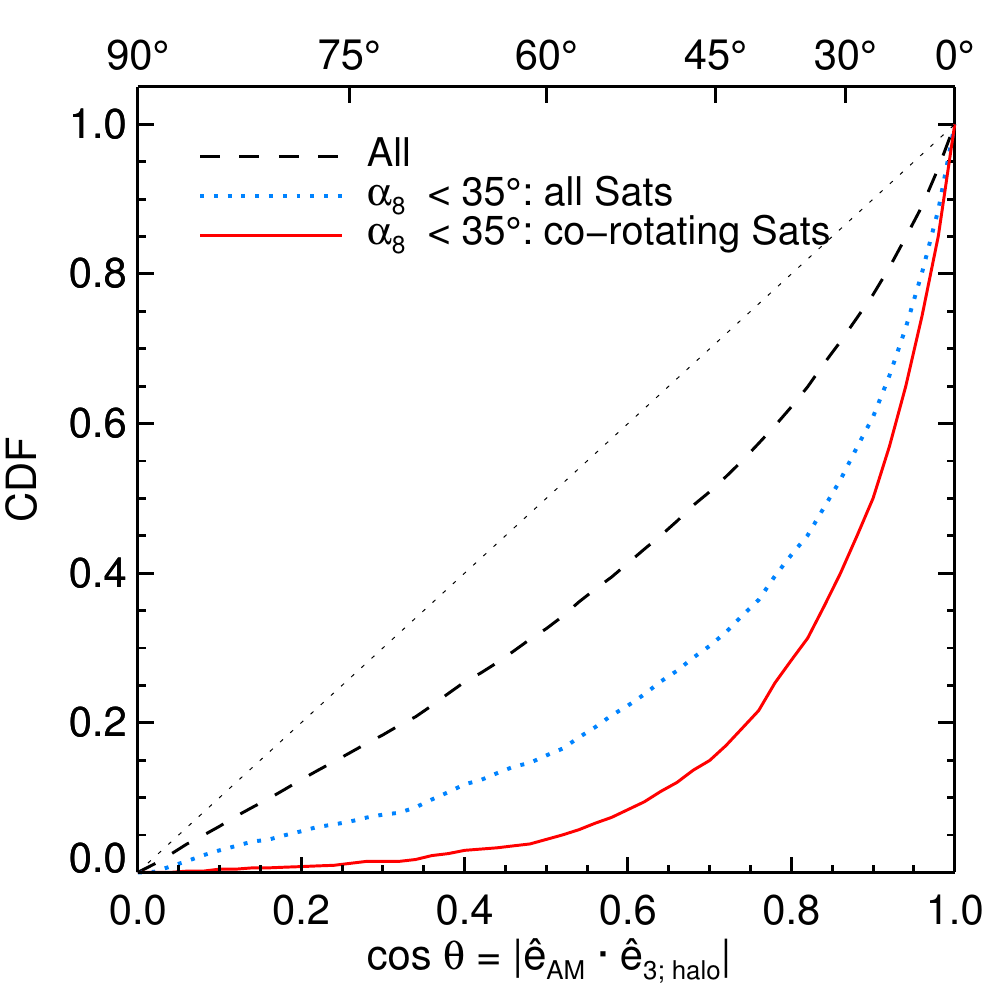}
	\caption{ The alignment of the satellite orbital poles with
          the minor axis of their host halo. The graph shows the CDF of the cosine of the 
          misalignment angle, $\cos \theta$, between
          the $z=0$ orbital angular momentum of satellites and the
          minor axis of their host halo. The dashed line shows the
          distribution for the brightest 11 satellites around all
          MW-mass hosts. The other two curves show results for
          \MWorbit{} systems (i.e. systems with
          $\alpha_8 < 35^\circ$); the dotted line is for all 11
          satellites of these systems, while the solid line is for 
          the 8 satellites with the most co-planar orbits. }
        \label{fig:Theta_op_halo}
\end{figure}
%%%%%%%%%%%%%%%%%%%%%%%%%%%%%%%%%%%%%%%%%%%%%%%%%%%%%%%%%%%%%%%%%%%%%
%%%%%%%%%%%%%%%%%%%%%%%%%%%%%%%%%%%%%%%%%%%%%%%%%%%%%%%%%%%%%%%%%%%%%
\begin{figure} \plotone{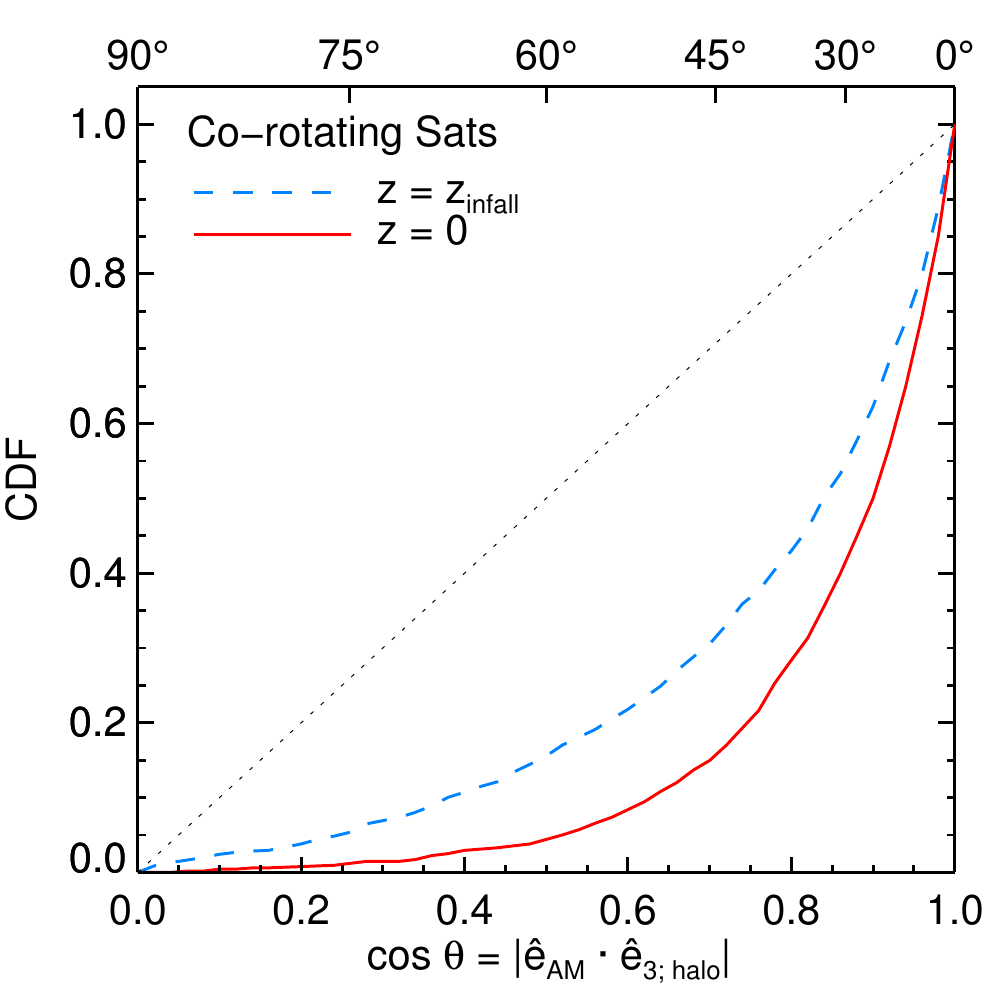} 
\caption{ The alignment of the satellites with co-planar orbits and the minor axis of
    their host halo. The graph shows the alignment of the angular momentum of
    the satellites at infall (dashed line) and of their
    present-day angular momentum (solid line). The stronger alignment
    for the $z=0$ orbital poles of co-planar satellites
    indicates that torques induced by the host halo are critical for
    the formation of rotating planes of satellites. }
	\label{fig:Theta_infall_halo}
\end{figure}
%%%%%%%%%%%%%%%%%%%%%%%%%%%%%%%%%%%%%%%%%%%%%%%%%%%%%%%%%%%%%%%%%%%%%

\subsubsection{The host halo as a driver of co-planar satellite orbits}
The host haloes of galactic satellites are triaxial and this gives
rise to torques that act upon the satellites. The effect of these
torques is complex and depends on the orientation of a satellite's
orbital pole with respect to its host halo \citep[see
e.g.][]{Bowden2013,Erkal2016,Fernando2017}. Orbital planes are
approximately stable in a triaxial halo only if they lie within the
equatorial or polar planes of the host halo. For all other
orientations, the orbit of a satellite can be thought of in terms of
box orbits and shows complex time variability \citep[see
e.g.][]{Pontzen2015}. These considerations suggest that the properties
of the host halo could play an important role in the formation and
survival of co-planar satellite distributions.

We investigate the role of the host halo in
\reffig{fig:Theta_op_halo}. This shows the cumulative distribution
function (CDF) of the alignment angle between the orbital poles of
satellites and the minor axis of their host halo. We can see that the
overall population of satellites shows mild alignment with the host
halo, with a preference for satellites to rotate within the equatorial
plane of their host \citep{Lovell2011,Cautun2015b}. The strength of
the alignment can be gauged by comparing with the dotted light grey
line, which shows the case of no alignment. The alignment between
satellite orbital poles and host halo is a manifestation of
anisotropic accretion as well as the torquing of orbits by the
aspherical mass distribution of the host.

Interestingly, the satellites of \MWorbit{} systems show a much
stronger alignment with their host haloes than the overall satellites
populations. To some extent, this would have been expected since
satellites in \MWorbit{} systems experience a greater degree of
anisotropic accretion (see \reffig{fig:Time_ca_infall}). However, the
difference in the alignment strength seems larger than the difference
in the distribution of $c/a$ values at infall (see
\reffig{fig:Time_ca_infall}), suggesting that additional physical
processes are at play. If we further restrict our analysis to the
satellites of \MWorbit{} systems that share the same orbital planes
(solid line in \reffig{fig:Theta_op_halo}), we find an even stronger
alignment between the orbital poles and the host's minor axis. This
suggests that co-planar satellites orbit preferentially in the
equatorial plane of their host halo. We study this topic in more
detail in a companion paper, \citet{Shao2019b}, where we show that in
the vast majority of cases, the normal to the plane in which the 8
co-planar satellite rotate is within $30^\circ$ of the halo minor
axis.

We now investigate if torques arising from the aspherical mass
distribution of the host halo contribute to the formation and
stability of co-rotating planes of satellites. On average, the torques
tend to move satellite orbits towards the host's equatorial plane and
thus this process would leave the following two signatures:
i)~co-rotating planes should lie close to the host halo's equatorial
plane, and ii)~after infall, the orbital plane of each co-planar
satellite should be systematically tilted towards the host's
equatorial plane. \reffig{fig:Theta_op_halo} shows that co-planar
satellites are well aligned with the host's minor axis and thus
provides evidence for the first signature of torques. However, a similar
alignment could result from the anisotropic accretion of satellites,
if the satellites entry points were found preferentially along the
host's equatorial plane \citep{Libeskind2014,Shao2018}. To prove
beyond doubt the role of the host halo, we need to check the second signature, which we do in
\reffig{fig:Theta_infall_halo}.

\reffig{fig:Theta_infall_halo} shows the change in a satellite's
orbital pole between infall and the present day, which we measure with
respect to the present-day host's minor axis. For clarity, we restrict
the analysis to the 8 co-planar satellites of \MWorbit{} systems. The
figure shows that the orbital poles of satellites are better aligned
with their host halo at the present than at infall, thus supporting
the hypothesis that tides resulting from the triaxial shape of the host
halo are important for the formation of rotating planes of satellite
galaxies. While not shown, we have performed other tests that support
the same conclusion. For example, the alignment between the orbital
poles of satellites with co-planar orbits and the minor axis of their
host halo always increases with time. Furthermore, the fraction of
\MWorbit{} systems increases with time from $7.3\%$ at $z=2$ to $11.3\%$
at the present day, which indicates that late-time processes play
an essential role in enhancing the number of satellites with co-planar orbits.

\section{Conclusions}
\label{sec:conclusions}
We have used the \eagle{} hydrocosmological simulation to study the
formation and evolution of planes of satellite galaxies similar to the
plane observed in the Milky Way. We have analysed the MW classical
dwarfs since these represent an observationally complete sample of
satellites that, moreover, are massive enough to be well resolved in
the \eagle{} simulation. Our sample consists of 1080 systems
with a typical halo mass of ${\sim}10^{12}\Msun$ that have at least 11
luminous satellites within a distance of $300~\rm{kpc}$ from the
centre.

To better understand the MW's satellite distribution, we have focused
on its two defining characteristics: its thinness and its high degree
of co-planar satellite orbits. To study thin planes of satellites we
have defined the \MWthin{} subsample which consists of \eagle{}
satellite systems with minor-to-major axes ratios, $c/a\leq0.3$. To
characterise co-planar orbits, we have devised a new, robust
formalism that identifies the subsets of satellites with the most
co-planar orbits. Out of the 11 MW classical satellites, 8 have highly
clustered orbital poles that are contained within a $21.9^\circ$
opening angle centred around $(l,b)=(182.3^\circ,-1.8^\circ)$. The
other 3 satellites -- Sextans, Leo I and Sagittarius -- have orbital
poles roughly perpendicular to this direction. The 8 Galactic
satellites with co-planar orbits stand out when compared to both
isotropic and to average $\Lambda$CDM satellite distributions. To
study these systems, we defined a \MWorbit{} subsample as those
systems for which at least 8 out of the brightest 11 satellites have
orbital poles contained within a $35^\circ$ opening angle.
In order to obtain large enough samples we deliberately adopt criteria
for identifying \textit{MW-like} planes that are less extreme than
the corresponding properties of the MW. Both our \MWthin{} and \MWorbit{}
subsamples were selected to contain ${\sim}100$ or more systems and
thus they allow us to infer statistically robust results. We expect
that similar processes to those that operate on our less extreme systems
will have also operated in the real MW.

\bigskip{}
\noindent ~~~~ Our main conclusions may be summarised as follows:
\\[-.45cm]
\begin{enumerate}
\item Most satellites have slowly-varying orbital planes; however
  there is a substantial population that experiences rapid and
  drastic changes in orbits. Most such events are due to close
  encounters with other massive satellites (see centre panels in
  \reffig{fig:MW_ca} and \ref{fig:examples}).
    
\item On average, satellite systems were flatter in the past, with a
  minimum $c/a$ axis ratio around $9~\rm{Gyrs}$ ago (see
  \reffig{fig:Time_ca}). This lookback time corresponds to the typical
  infall time of the classical satellites and suggests that the
  progenitors of the present-day satellites first collapsed onto a
  large-scale sheet before falling into their host haloes. After
  infall, satellite systems are characterised by increasingly higher
  $c/a$ ratios (see \reffig{fig:Time_ca_infall}).
    
\item \MWthin{} planes of satellites are short-lived chance
  associations whose thickness increases on a timescale of less
  than $1~\rm{Gyr}$. However, such systems are at all times 
  flatter than the overall satellite system population (see
  \reffig{fig:Time_ca}).
    
\item Both \MWthin{} and \MWorbit{} systems were, on average, flatter at
  infall than the overall population of galactic satellites (see
  \reffig{fig:Time_ca_infall}). However, a highly anisotropic
  satellite infall does not guarantee the formation of a plane of
  satellites, indicating that anisotropic accretion is just one
  of the processes that result in the formation of planes of
  satellites.
    
\item We find a poor correlation between the direction of planes of
  satellites at two different epochs. The correlation is stronger for
  \MWorbit{} systems where many satellites rotate in the same plane,
  but even in this case, the misalignment angle is $\geq 35^\circ$
  after just $2~\rm{Gyrs}$ (see \reffig{fig:Time_e3}).
    
\item Satellites that orbit within the same plane have not always done
  so and around half of them formed a plane within the last
  ${\sim}4~\rm{Gyrs}$. There is a large system-to-system variation, 
  with significant fractions forming both recently and as long as
  $9 ~\rm{Gyrs}$ ago (see \reffig{fig:Time_a8}).
    
\item Co-planar satellites orbit preferentially within the
  equatorial plane of their host halo (see
  \reffig{fig:Theta_op_halo}). The torques resulting from the
  aspherical mass distribution of the hosts play an important
  role in the formation of rotating planes of satellites (see
  \reffig{fig:Theta_infall_halo}). \end{enumerate}

\bigskip{} 
One of the goals of this paper has been to understand the
Galactic plane of classical satellites. Our results indicate that the
flattening of the MW satellite distribution is short lived \citep[see
also][]{Lipnicky2017} and likely due to a chance occurrence. In
contrast, the large number of satellites, 8 out of the 11 classical
dwarfs, that share the same orbital plane is physically more
interesting. We predict that a majority of these 8 satellites have
been orbiting within the same plane for many billions of years. In
particular, the \eagle{} simulation predicts that a smaller opening
angle enclosing the orbital poles of the 8 co-planar satellites
corresponds to an older rotating plane of satellites. The MW opening
angle lies in the tail of the distribution and thus indicates a very
long-lived rotating plane composed of 8 out of the 11 classical
satellites. Note, however, that at least two of its members, the LMC
and the SMC, are thought to be recent additions to the MW satellite
system \citep{Besla2007,Kallivayalil2013,Patel2017}. For these two
galaxies, we predict that their orbital angular momentum has pointed
roughly in the same direction since long before their infall into the
MW.

From a practical perspective, the 8 satellites with co-planar orbits
can be used to infer, with high precision, the orientation of the
Galactic dark matter halo, as we show in a companion paper
\citep{Shao2019b}. Furthermore, the halo tidal field plays an
important role in the formation of co-planar satellites and better
understanding this process could lead to additional inferences about
the Galactic dark matter halo and its relation to the cosmic web
around our galaxy.

While we infer an early formation time for the Galactic subset of the
8 satellites with co-planar orbits, its future could be short-lived.
This is a consequence of the recent accretion of the LMC, whose
massive dark matter halo \citep{Penarrubia2016,Shao2018b} can induce
dramatic changes in the orbital planes of the other dwarfs. This is
especially true if close encounters occur; observational data already
shows that this likely happened recently for the Tucana III stream
\citep{Erkal2018}. Luckily, the LMC is predicted to merge with our
galaxy in the next ${\sim}2.5~\rm{Gyrs}$ \citep{Cautun2019}; this may
allow the subset of co-planar satellites to survive unscathed, albeit
with at least one fewer member.

 \vspace{-.3cm}
\section*{Acknowledgements}
We thank the anonymous referee for detailed comments that have helped us improve the paper.
SS, MC and CSF were supported by the Science and Technology Facilities Council (STFC) [grant number ST/I00162X/1, ST/P000541/1] and by the ERC Advanced Investigator grant, DMIDAS [GA 786910]. SS is also supported by the ERC grant ERC-StG-716532-PUNCA.
This work used the DiRAC@Durham facility managed by the Institute for
Computational Cosmology on behalf of the STFC DiRAC HPC Facility
(www.dirac.ac.uk). The equipment was funded by BEIS capital funding
via STFC capital grants ST/K00042X/1, ST/P002293/1, ST/R002371/1 and
ST/S002502/1, Durham University and STFC operations grant
ST/R000832/1. DiRAC is part of the National e-Infrastructure.

\vspace{-.3cm}
\bibliographystyle{mnras}
\bibliography{bibliography}

\begin{thebibliography}{}
\makeatletter
\relax
\def\mn@urlcharsother{\let\do\@makeother \do\$\do\&\do\#\do\^\do\_\do\%\do\~}
\def\mn@doi{\begingroup\mn@urlcharsother \@ifnextchar [ {\mn@doi@}
  {\mn@doi@[]}}
\def\mn@doi@[#1]#2{\def\@tempa{#1}\ifx\@tempa\@empty \href
  {http://dx.doi.org/#2} {doi:#2}\else \href {http://dx.doi.org/#2} {#1}\fi
  \endgroup}
\def\mn@eprint#1#2{\mn@eprint@#1:#2::\@nil}
\def\mn@eprint@arXiv#1{\href {http://arxiv.org/abs/#1} {{\tt arXiv:#1}}}
\def\mn@eprint@dblp#1{\href {http://dblp.uni-trier.de/rec/bibtex/#1.xml}
  {dblp:#1}}
\def\mn@eprint@#1:#2:#3:#4\@nil{\def\@tempa {#1}\def\@tempb {#2}\def\@tempc
  {#3}\ifx \@tempc \@empty \let \@tempc \@tempb \let \@tempb \@tempa \fi \ifx
  \@tempb \@empty \def\@tempb {arXiv}\fi \@ifundefined
  {mn@eprint@\@tempb}{\@tempb:\@tempc}{\expandafter \expandafter \csname
  mn@eprint@\@tempb\endcsname \expandafter{\@tempc}}}

\bibitem[\protect\citeauthoryear{{Agustsson} \& {Brainerd}}{{Agustsson} \&
  {Brainerd}}{2010}]{Agustsson2010}
{Agustsson} I.,  {Brainerd} T.~G.,  2010, \mn@doi [\apj]
  {10.1088/0004-637X/709/2/1321}, \href
  {http://adsabs.harvard.edu/abs/2010ApJ...709.1321A} {709, 1321}

\bibitem[\protect\citeauthoryear{{Ahmed}, {Brooks}  \& {Christensen}}{{Ahmed}
  et~al.}{2017}]{Ahmed2017}
{Ahmed} S.~H.,  {Brooks} A.~M.,   {Christensen} C.~R.,  2017, \mn@doi [\mnras]
  {10.1093/mnras/stw3271}, \href
  {http://adsabs.harvard.edu/abs/2017MNRAS.466.3119A} {466, 3119}

\bibitem[\protect\citeauthoryear{{Arnold}, {Shandarin}  \&
  {Zeldovich}}{{Arnold} et~al.}{1982}]{Arnold1982}
{Arnold} V.~I.,  {Shandarin} S.~F.,   {Zeldovich} I.~B.,  1982, \mn@doi
  [Geophysical and Astrophysical Fluid Dynamics] {10.1080/03091928208209001},
  \href {http://adsabs.harvard.edu/abs/1982GApFD..20..111A} {20, 111}

\bibitem[\protect\citeauthoryear{{Aubert}, {Pichon}  \& {Colombi}}{{Aubert}
  et~al.}{2004}]{Aubert2004}
{Aubert} D.,  {Pichon} C.,   {Colombi} S.,  2004, \mn@doi [\mnras]
  {10.1111/j.1365-2966.2004.07883.x}, \href
  {http://adsabs.harvard.edu/abs/2004MNRAS.352..376A} {352, 376}

\bibitem[\protect\citeauthoryear{{Bahl} \& {Baumgardt}}{{Bahl} \&
  {Baumgardt}}{2014}]{Bahl2014}
{Bahl} H.,  {Baumgardt} H.,  2014, \mn@doi [\mnras] {10.1093/mnras/stt2399},
  \href {http://adsabs.harvard.edu/abs/2014MNRAS.438.2916B} {438, 2916}

\bibitem[\protect\citeauthoryear{{Banik}, {O'Ryan}  \& {Zhao}}{{Banik}
  et~al.}{2018}]{Banik2018}
{Banik} I.,  {O'Ryan} D.,   {Zhao} H.,  2018, \mn@doi [\mnras]
  {10.1093/mnras/sty919}, \href
  {http://adsabs.harvard.edu/abs/2018MNRAS.477.4768B} {477, 4768}

\bibitem[\protect\citeauthoryear{{Besla}, {Kallivayalil}, {Hernquist},
  {Robertson}, {Cox}, {van der Marel}  \& {Alcock}}{{Besla}
  et~al.}{2007}]{Besla2007}
{Besla} G.,  {Kallivayalil} N.,  {Hernquist} L.,  {Robertson} B.,  {Cox} T.~J.,
   {van der Marel} R.~P.,   {Alcock} C.,  2007, \mn@doi [\apj]
  {10.1086/521385}, \href {http://adsabs.harvard.edu/abs/2007ApJ...668..949B}
  {668, 949}

\bibitem[\protect\citeauthoryear{{Bowden}, {Evans}  \& {Belokurov}}{{Bowden}
  et~al.}{2013}]{Bowden2013}
{Bowden} A.,  {Evans} N.~W.,   {Belokurov} V.,  2013, \mn@doi [\mnras]
  {10.1093/mnras/stt1253}, \href
  {http://adsabs.harvard.edu/abs/2013MNRAS.435..928B} {435, 928}

\bibitem[\protect\citeauthoryear{{Brainerd}}{{Brainerd}}{2005}]{Brainerd2005}
{Brainerd} T.~G.,  2005, \mn@doi [\apjl] {10.1086/432713}, \href
  {http://adsabs.harvard.edu/abs/2005ApJ...628L.101B} {628, L101}

\bibitem[\protect\citeauthoryear{{Buck}, {Macci{\`o}}  \& {Dutton}}{{Buck}
  et~al.}{2015}]{Buck2015}
{Buck} T.,  {Macci{\`o}} A.~V.,   {Dutton} A.~A.,  2015, \mn@doi [\apj]
  {10.1088/0004-637X/809/1/49}, \href
  {http://adsabs.harvard.edu/abs/2015ApJ...809...49B} {809, 49}

\bibitem[\protect\citeauthoryear{{Buck}, {Dutton}  \& {Macci{\`o}}}{{Buck}
  et~al.}{2016}]{Buck2016}
{Buck} T.,  {Dutton} A.~A.,   {Macci{\`o}} A.~V.,  2016, \mn@doi [\mnras]
  {10.1093/mnras/stw1232}, \href
  {http://adsabs.harvard.edu/abs/2016MNRAS.460.4348B} {460, 4348}

\bibitem[\protect\citeauthoryear{{Callingham} et~al.,}{{Callingham}
  et~al.}{2019}]{Callingham2018}
{Callingham} T.~M.,  et~al., 2019, \mn@doi [\mnras] {10.1093/mnras/stz365},
  \href {https://ui.adsabs.harvard.edu/abs/2019MNRAS.484.5453C} {484, 5453}

\bibitem[\protect\citeauthoryear{{Cautun} \& {Frenk}}{{Cautun} \&
  {Frenk}}{2017}]{Cautun2017}
{Cautun} M.,  {Frenk} C.~S.,  2017, \mn@doi [\mnras] {10.1093/mnrasl/slx025},
  \href {http://adsabs.harvard.edu/abs/2017MNRAS.468L..41C} {468, L41}

\bibitem[\protect\citeauthoryear{{Cautun}, {van de Weygaert}, {Jones}  \&
  {Frenk}}{{Cautun} et~al.}{2014}]{Cautun2014c}
{Cautun} M.,  {van de Weygaert} R.,  {Jones} B.~J.~T.,   {Frenk} C.~S.,  2014,
  \mn@doi [\mnras] {10.1093/mnras/stu768}, \href
  {http://adsabs.harvard.edu/abs/2014MNRAS.441.2923C} {441, 2923}

\bibitem[\protect\citeauthoryear{{Cautun}, {Wang}, {Frenk}  \&
  {Sawala}}{{Cautun} et~al.}{2015a}]{Cautun2015b}
{Cautun} M.,  {Wang} W.,  {Frenk} C.~S.,   {Sawala} T.,  2015a, \mn@doi
  [\mnras] {10.1093/mnras/stv490}, \href
  {http://adsabs.harvard.edu/abs/2015MNRAS.449.2576C} {449, 2576}

\bibitem[\protect\citeauthoryear{{Cautun}, {Bose}, {Frenk}, {Guo}, {Han},
  {Hellwing}, {Sawala}  \& {Wang}}{{Cautun} et~al.}{2015b}]{Cautun2015}
{Cautun} M.,  {Bose} S.,  {Frenk} C.~S.,  {Guo} Q.,  {Han} J.,  {Hellwing}
  W.~A.,  {Sawala} T.,   {Wang} W.,  2015b, \mn@doi [\mnras]
  {10.1093/mnras/stv1557}, \href
  {http://adsabs.harvard.edu/abs/2015MNRAS.452.3838C} {452, 3838}

\bibitem[\protect\citeauthoryear{{Cautun}, {Deason}, {Frenk}  \&
  {McAlpine}}{{Cautun} et~al.}{2019}]{Cautun2019}
{Cautun} M.,  {Deason} A.~J.,  {Frenk} C.~S.,   {McAlpine} S.,  2019, \mn@doi
  [\mnras] {10.1093/mnras/sty3084}, \href
  {https://ui.adsabs.harvard.edu/\#abs/2019MNRAS.483.2185C} {483, 2185}

\bibitem[\protect\citeauthoryear{{Conn} et~al.,}{{Conn}
  et~al.}{2013}]{Conn2013}
{Conn} A.~R.,  et~al., 2013, \mn@doi [\apj] {10.1088/0004-637X/766/2/120},
  \href {http://adsabs.harvard.edu/abs/2013ApJ...766..120C} {766, 120}

\bibitem[\protect\citeauthoryear{{Crain} et~al.,}{{Crain}
  et~al.}{2015}]{Crain2015}
{Crain} R.~A.,  et~al., 2015, \mn@doi [\mnras] {10.1093/mnras/stv725}, \href
  {http://adsabs.harvard.edu/abs/2015MNRAS.450.1937C} {450, 1937}

\bibitem[\protect\citeauthoryear{{Danovich}, {Dekel}, {Hahn}  \&
  {Teyssier}}{{Danovich} et~al.}{2012}]{Danovich2012}
{Danovich} M.,  {Dekel} A.,  {Hahn} O.,   {Teyssier} R.,  2012, \mn@doi
  [\mnras] {10.1111/j.1365-2966.2012.20751.x}, \href
  {http://adsabs.harvard.edu/abs/2012MNRAS.422.1732D} {422, 1732}

\bibitem[\protect\citeauthoryear{{Davis}, {Efstathiou}, {Frenk}  \&
  {White}}{{Davis} et~al.}{1985}]{Davis1985}
{Davis} M.,  {Efstathiou} G.,  {Frenk} C.~S.,   {White} S.~D.~M.,  1985,
  \mn@doi [\apj] {10.1086/163168}, \href
  {http://adsabs.harvard.edu/abs/1985ApJ...292..371D} {292, 371}

\bibitem[\protect\citeauthoryear{{Deason} et~al.,}{{Deason}
  et~al.}{2011}]{Deason2011}
{Deason} A.~J.,  et~al., 2011, \mn@doi [\mnras]
  {10.1111/j.1365-2966.2011.18884.x}, \href
  {http://adsabs.harvard.edu/abs/2011MNRAS.415.2607D} {415, 2607}

\bibitem[\protect\citeauthoryear{{Dolag}, {Borgani}, {Murante}  \&
  {Springel}}{{Dolag} et~al.}{2009}]{Dolag2009}
{Dolag} K.,  {Borgani} S.,  {Murante} G.,   {Springel} V.,  2009, \mn@doi
  [\mnras] {10.1111/j.1365-2966.2009.15034.x}, \href
  {http://adsabs.harvard.edu/abs/2009MNRAS.399..497D} {399, 497}

\bibitem[\protect\citeauthoryear{{Erkal}, {Sanders}  \& {Belokurov}}{{Erkal}
  et~al.}{2016}]{Erkal2016}
{Erkal} D.,  {Sanders} J.~L.,   {Belokurov} V.,  2016, \mn@doi [\mnras]
  {10.1093/mnras/stw1400}, \href
  {http://adsabs.harvard.edu/abs/2016MNRAS.461.1590E} {461, 1590}

\bibitem[\protect\citeauthoryear{{Erkal} et~al.,}{{Erkal}
  et~al.}{2018}]{Erkal2018}
{Erkal} D.,  et~al., 2018, \mn@doi [\mnras] {10.1093/mnras/sty2518}, \href
  {http://adsabs.harvard.edu/abs/2018MNRAS.481.3148E} {481, 3148}

\bibitem[\protect\citeauthoryear{{Fernando}, {Arias}, {Guglielmo}, {Lewis},
  {Ibata}  \& {Power}}{{Fernando} et~al.}{2017}]{Fernando2017}
{Fernando} N.,  {Arias} V.,  {Guglielmo} M.,  {Lewis} G.~F.,  {Ibata} R.~A.,
  {Power} C.,  2017, \mn@doi [\mnras] {10.1093/mnras/stw2694}, \href
  {http://adsabs.harvard.edu/abs/2017MNRAS.465..641F} {465, 641}

\bibitem[\protect\citeauthoryear{{Fernando}, {Arias}, {Lewis}, {Ibata}  \&
  {Power}}{{Fernando} et~al.}{2018}]{Fernando2018}
{Fernando} N.,  {Arias} V.,  {Lewis} G.~F.,  {Ibata} R.~A.,   {Power} C.,
  2018, \mn@doi [\mnras] {10.1093/mnras/stx2483}, \href
  {http://adsabs.harvard.edu/abs/2018MNRAS.473.2212F} {473, 2212}

\bibitem[\protect\citeauthoryear{{Fritz}, {Battaglia}, {Pawlowski},
  {Kallivayalil}, {van der Marel}, {Sohn}, {Brook}  \& {Besla}}{{Fritz}
  et~al.}{2018}]{Fritz2018}
{Fritz} T.~K.,  {Battaglia} G.,  {Pawlowski} M.~S.,  {Kallivayalil} N.,  {van
  der Marel} R.,  {Sohn} S.~T.,  {Brook} C.,   {Besla} G.,  2018, \mn@doi
  [\aap] {10.1051/0004-6361/201833343}, \href
  {http://adsabs.harvard.edu/abs/2018A%26A...619A.103F} {619, A103}

\bibitem[\protect\citeauthoryear{{Gaia Collaboration} et~al.,}{{Gaia
  Collaboration} et~al.}{2018}]{Gaia2018}
{Gaia Collaboration} et~al., 2018, \mn@doi [\aap]
  {10.1051/0004-6361/201832698}, \href
  {http://adsabs.harvard.edu/abs/2018A%26A...616A..12G} {616, A12}

\bibitem[\protect\citeauthoryear{{Gillet}, {Ocvirk}, {Aubert}, {Knebe},
  {Libeskind}, {Yepes}, {Gottl{\"o}ber}  \& {Hoffman}}{{Gillet}
  et~al.}{2015}]{Gillet2015}
{Gillet} N.,  {Ocvirk} P.,  {Aubert} D.,  {Knebe} A.,  {Libeskind} N.,  {Yepes}
  G.,  {Gottl{\"o}ber} S.,   {Hoffman} Y.,  2015, \mn@doi [\apj]
  {10.1088/0004-637X/800/1/34}, \href
  {https://ui.adsabs.harvard.edu/\#abs/2015ApJ...800...34G} {800, 34}

\bibitem[\protect\citeauthoryear{{G{\'o}mez}, {Besla}, {Carpintero},
  {Villalobos}, {O'Shea}  \& {Bell}}{{G{\'o}mez} et~al.}{2015}]{Gomez2015}
{G{\'o}mez} F.~A.,  {Besla} G.,  {Carpintero} D.~D.,  {Villalobos} {\'A}.,
  {O'Shea} B.~W.,   {Bell} E.~F.,  2015, \mn@doi [\apj]
  {10.1088/0004-637X/802/2/128}, \href
  {http://adsabs.harvard.edu/abs/2015ApJ...802..128G} {802, 128}

\bibitem[\protect\citeauthoryear{{Gonz{\'a}lez} \& {Padilla}}{{Gonz{\'a}lez} \&
  {Padilla}}{2016}]{Gonzalez2016}
{Gonz{\'a}lez} R.~E.,  {Padilla} N.~D.,  2016, \mn@doi [\apj]
  {10.3847/0004-637X/829/1/58}, \href
  {http://adsabs.harvard.edu/abs/2016ApJ...829...58G} {829, 58}

\bibitem[\protect\citeauthoryear{{Hammer}, {Yang}, {Fouquet}, {Pawlowski},
  {Kroupa}, {Puech}, {Flores}  \& {Wang}}{{Hammer} et~al.}{2013}]{Hammer2013}
{Hammer} F.,  {Yang} Y.,  {Fouquet} S.,  {Pawlowski} M.~S.,  {Kroupa} P.,
  {Puech} M.,  {Flores} H.,   {Wang} J.,  2013, \mn@doi [\mnras]
  {10.1093/mnras/stt435}, \href
  {http://adsabs.harvard.edu/abs/2013MNRAS.431.3543H} {431, 3543}

\bibitem[\protect\citeauthoryear{{Hodkinson} \& {Scholtz}}{{Hodkinson} \&
  {Scholtz}}{2019}]{Hodkinson2019}
{Hodkinson} B.,  {Scholtz} J.,  2019, arXiv e-prints, \href
  {https://ui.adsabs.harvard.edu/abs/2019arXiv190403192H} {}

\bibitem[\protect\citeauthoryear{{Ibata} et~al.,}{{Ibata}
  et~al.}{2013}]{Ibata2013}
{Ibata} R.~A.,  et~al., 2013, \mn@doi [\nat] {10.1038/nature11717}, \href
  {http://adsabs.harvard.edu/abs/2013Natur.493...62I} {493, 62}

\bibitem[\protect\citeauthoryear{{Icke}}{{Icke}}{1973}]{Icke1973}
{Icke} V.,  1973, \aap, \href
  {http://adsabs.harvard.edu/abs/1973A%26A....27....1I} {27, 1}

\bibitem[\protect\citeauthoryear{{Jiang}, {Helly}, {Cole}  \& {Frenk}}{{Jiang}
  et~al.}{2014}]{Jiang2014}
{Jiang} L.,  {Helly} J.~C.,  {Cole} S.,   {Frenk} C.~S.,  2014, \mn@doi
  [\mnras] {10.1093/mnras/stu390}, \href
  {http://adsabs.harvard.edu/abs/2014MNRAS.440.2115J} {440, 2115}

\bibitem[\protect\citeauthoryear{{Kallivayalil}, {van der Marel}, {Besla},
  {Anderson}  \& {Alcock}}{{Kallivayalil} et~al.}{2013}]{Kallivayalil2013}
{Kallivayalil} N.,  {van der Marel} R.~P.,  {Besla} G.,  {Anderson} J.,
  {Alcock} C.,  2013, \mn@doi [\apj] {10.1088/0004-637X/764/2/161}, \href
  {http://adsabs.harvard.edu/abs/2013ApJ...764..161K} {764, 161}

\bibitem[\protect\citeauthoryear{{Kang}, {Mao}, {Gao}  \& {Jing}}{{Kang}
  et~al.}{2005}]{Kang2005}
{Kang} X.,  {Mao} S.,  {Gao} L.,   {Jing} Y.~P.,  2005, \mn@doi [\aap]
  {10.1051/0004-6361:20052675}, \href
  {http://adsabs.harvard.edu/abs/2005A%26A...437..383K} {437, 383}

\bibitem[\protect\citeauthoryear{{Knebe}, {Gill}, {Gibson}, {Lewis}, {Ibata}
  \& {Dopita}}{{Knebe} et~al.}{2004}]{Knebe2004}
{Knebe} A.,  {Gill} S.~P.~D.,  {Gibson} B.~K.,  {Lewis} G.~F.,  {Ibata} R.~A.,
   {Dopita} M.~A.,  2004, \mn@doi [\apj] {10.1086/381306}, \href
  {http://adsabs.harvard.edu/abs/2004ApJ...603....7K} {603, 7}

\bibitem[\protect\citeauthoryear{{Kroupa}, {Theis}  \& {Boily}}{{Kroupa}
  et~al.}{2005}]{Kroupa2005}
{Kroupa} P.,  {Theis} C.,   {Boily} C.~M.,  2005, \mn@doi [\aap]
  {10.1051/0004-6361:20041122}, \href
  {http://adsabs.harvard.edu/abs/2005A%26A...431..517K} {431, 517}

\bibitem[\protect\citeauthoryear{{Kunkel} \& {Demers}}{{Kunkel} \&
  {Demers}}{1976}]{Kunkel1976}
{Kunkel} W.~E.,  {Demers} S.,  1976, in {Dickens} R.~J.,  {Perry} J.~E.,
  {Smith} F.~G.,   {King} I.~R.,  eds,  Royal Greenwich Observatory Bulletins
  Vol. 182, The Galaxy and the Local Group. p.~241

\bibitem[\protect\citeauthoryear{{Li} \& {Helmi}}{{Li} \&
  {Helmi}}{2008}]{Li2008}
{Li} Y.-S.,  {Helmi} A.,  2008, \mn@doi [\mnras]
  {10.1111/j.1365-2966.2008.12854.x}, \href
  {http://adsabs.harvard.edu/abs/2008MNRAS.385.1365L} {385, 1365}

\bibitem[\protect\citeauthoryear{{Libeskind}, {Frenk}, {Cole}, {Helly},
  {Jenkins}, {Navarro}  \& {Power}}{{Libeskind} et~al.}{2005}]{Libeskind2005}
{Libeskind} N.~I.,  {Frenk} C.~S.,  {Cole} S.,  {Helly} J.~C.,  {Jenkins} A.,
  {Navarro} J.~F.,   {Power} C.,  2005, \mn@doi [\mnras]
  {10.1111/j.1365-2966.2005.09425.x}, \href
  {http://adsabs.harvard.edu/abs/2005MNRAS.363..146L} {363, 146}

\bibitem[\protect\citeauthoryear{{Libeskind}, {Cole}, {Frenk}, {Okamoto}  \&
  {Jenkins}}{{Libeskind} et~al.}{2007}]{Libeskind2007}
{Libeskind} N.~I.,  {Cole} S.,  {Frenk} C.~S.,  {Okamoto} T.,   {Jenkins} A.,
  2007, \mn@doi [\mnras] {10.1111/j.1365-2966.2006.11205.x}, \href
  {http://adsabs.harvard.edu/abs/2007MNRAS.374...16L} {374, 16}

\bibitem[\protect\citeauthoryear{{Libeskind}, {Frenk}, {Cole}, {Jenkins}  \&
  {Helly}}{{Libeskind} et~al.}{2009}]{Libeskind2009}
{Libeskind} N.~I.,  {Frenk} C.~S.,  {Cole} S.,  {Jenkins} A.,   {Helly} J.~C.,
  2009, \mn@doi [\mnras] {10.1111/j.1365-2966.2009.15315.x}, \href
  {http://adsabs.harvard.edu/abs/2009MNRAS.399..550L} {399, 550}

\bibitem[\protect\citeauthoryear{{Libeskind}, {Knebe}, {Hoffman}  \&
  {Gottl{\"o}ber}}{{Libeskind} et~al.}{2014}]{Libeskind2014}
{Libeskind} N.~I.,  {Knebe} A.,  {Hoffman} Y.,   {Gottl{\"o}ber} S.,  2014,
  \mn@doi [\mnras] {10.1093/mnras/stu1216}, \href
  {http://adsabs.harvard.edu/abs/2014MNRAS.443.1274L} {443, 1274}

\bibitem[\protect\citeauthoryear{{Lipnicky} \& {Chakrabarti}}{{Lipnicky} \&
  {Chakrabarti}}{2017}]{Lipnicky2017}
{Lipnicky} A.,  {Chakrabarti} S.,  2017, \mn@doi [\mnras]
  {10.1093/mnras/stx286}, \href
  {http://adsabs.harvard.edu/abs/2017MNRAS.468.1671L} {468, 1671}

\bibitem[\protect\citeauthoryear{{Lovell}, {Eke}, {Frenk}  \&
  {Jenkins}}{{Lovell} et~al.}{2011}]{Lovell2011}
{Lovell} M.~R.,  {Eke} V.~R.,  {Frenk} C.~S.,   {Jenkins} A.,  2011, \mn@doi
  [\mnras] {10.1111/j.1365-2966.2011.18377.x}, \href
  {http://adsabs.harvard.edu/abs/2011MNRAS.413.3013L} {413, 3013}

\bibitem[\protect\citeauthoryear{{Lynden-Bell}}{{Lynden-Bell}}{1976}]{Lynden-Bell1976}
{Lynden-Bell} D.,  1976, \mnras, \href
  {http://adsabs.harvard.edu/abs/1976MNRAS.174..695L} {174, 695}

\bibitem[\protect\citeauthoryear{{Lynden-Bell}}{{Lynden-Bell}}{1982}]{Lynden-Bell1982}
{Lynden-Bell} D.,  1982, The Observatory, \href
  {http://adsabs.harvard.edu/abs/1982Obs...102..202L} {102, 202}

\bibitem[\protect\citeauthoryear{{McAlpine} et~al.,}{{McAlpine}
  et~al.}{2016}]{McAlpine2016}
{McAlpine} S.,  et~al., 2016, \mn@doi [Astronomy and Computing]
  {10.1016/j.ascom.2016.02.004}, \href
  {http://adsabs.harvard.edu/abs/2016A%26C....15...72M} {15, 72}

\bibitem[\protect\citeauthoryear{{McConnachie}}{{McConnachie}}{2012}]{McConnachie2012}
{McConnachie} A.~W.,  2012, \mn@doi [\aj] {10.1088/0004-6256/144/1/4}, \href
  {http://adsabs.harvard.edu/abs/2012AJ....144....4M} {144, 4}

\bibitem[\protect\citeauthoryear{{Metz}, {Kroupa}  \& {Libeskind}}{{Metz}
  et~al.}{2008}]{Metz2008}
{Metz} M.,  {Kroupa} P.,   {Libeskind} N.~I.,  2008, \mn@doi [\apj]
  {10.1086/587833}, \href {http://adsabs.harvard.edu/abs/2008ApJ...680..287M}
  {680, 287}

\bibitem[\protect\citeauthoryear{{Metz}, {Kroupa}, {Theis}, {Hensler}  \&
  {Jerjen}}{{Metz} et~al.}{2009}]{Metz2009}
{Metz} M.,  {Kroupa} P.,  {Theis} C.,  {Hensler} G.,   {Jerjen} H.,  2009,
  \mn@doi [\apj] {10.1088/0004-637X/697/1/269}, \href
  {http://adsabs.harvard.edu/abs/2009ApJ...697..269M} {697, 269}

\bibitem[\protect\citeauthoryear{{M{\"u}ller}, {Pawlowski}, {Jerjen}  \&
  {Lelli}}{{M{\"u}ller} et~al.}{2018}]{Muller2018}
{M{\"u}ller} O.,  {Pawlowski} M.~S.,  {Jerjen} H.,   {Lelli} F.,  2018, \mn@doi
  [Science] {10.1126/science.aao1858}, \href
  {http://adsabs.harvard.edu/abs/2018Sci...359..534M} {359, 534}

\bibitem[\protect\citeauthoryear{{Newton}, {Cautun}, {Jenkins}, {Frenk}  \&
  {Helly}}{{Newton} et~al.}{2018}]{Newton2018}
{Newton} O.,  {Cautun} M.,  {Jenkins} A.,  {Frenk} C.~S.,   {Helly} J.~C.,
  2018, \mn@doi [\mnras] {10.1093/mnras/sty1085}, \href
  {http://adsabs.harvard.edu/abs/2018MNRAS.479.2853N} {479, 2853}

\bibitem[\protect\citeauthoryear{{Nierenberg}, {Auger}, {Treu}, {Marshall},
  {Fassnacht}  \& {Busha}}{{Nierenberg} et~al.}{2012}]{Nierenberg2012}
{Nierenberg} A.~M.,  {Auger} M.~W.,  {Treu} T.,  {Marshall} P.~J.,  {Fassnacht}
  C.~D.,   {Busha} M.~T.,  2012, \mn@doi [\apj] {10.1088/0004-637X/752/2/99},
  \href {http://adsabs.harvard.edu/abs/2012ApJ...752...99N} {752, 99}

\bibitem[\protect\citeauthoryear{{Patel}, {Besla}  \& {Sohn}}{{Patel}
  et~al.}{2017}]{Patel2017}
{Patel} E.,  {Besla} G.,   {Sohn} S.~T.,  2017, \mn@doi [\mnras]
  {10.1093/mnras/stw2616}, \href
  {http://adsabs.harvard.edu/abs/2017MNRAS.464.3825P} {464, 3825}

\bibitem[\protect\citeauthoryear{{Pawlowski}}{{Pawlowski}}{2016}]{Pawlowski2016}
{Pawlowski} M.~S.,  2016, \mn@doi [\mnras] {10.1093/mnras/stv2673}, \href
  {http://adsabs.harvard.edu/abs/2016MNRAS.456..448P} {456, 448}

\bibitem[\protect\citeauthoryear{{Pawlowski} \& {Kroupa}}{{Pawlowski} \&
  {Kroupa}}{2013}]{Pawlowski2013b}
{Pawlowski} M.~S.,  {Kroupa} P.,  2013, \mn@doi [\mnras]
  {10.1093/mnras/stt1429}, \href
  {http://adsabs.harvard.edu/abs/2013MNRAS.435.2116P} {435, 2116}

\bibitem[\protect\citeauthoryear{{Pawlowski}, {Kroupa}, {Angus}, {de Boer},
  {Famaey}  \& {Hensler}}{{Pawlowski} et~al.}{2012}]{Pawlowski2012b}
{Pawlowski} M.~S.,  {Kroupa} P.,  {Angus} G.,  {de Boer} K.~S.,  {Famaey} B.,
  {Hensler} G.,  2012, \mn@doi [\mnras] {10.1111/j.1365-2966.2012.21169.x},
  \href {http://adsabs.harvard.edu/abs/2012MNRAS.424...80P} {424, 80}

\bibitem[\protect\citeauthoryear{{Pawlowski} et~al.,}{{Pawlowski}
  et~al.}{2014}]{Pawlowski2014c}
{Pawlowski} M.~S.,  et~al., 2014, \mn@doi [\mnras] {10.1093/mnras/stu1005},
  \href {http://adsabs.harvard.edu/abs/2014MNRAS.442.2362P} {442, 2362}

\bibitem[\protect\citeauthoryear{{Pawlowski} et~al.,}{{Pawlowski}
  et~al.}{2017}]{Pawlowski2017AN}
{Pawlowski} M.~S.,  et~al., 2017, \mn@doi [Astronomische Nachrichten]
  {10.1002/asna.201713366}, \href
  {https://ui.adsabs.harvard.edu/abs/2017AN....338..854P} {338, 854}

\bibitem[\protect\citeauthoryear{{Pe{\~n}arrubia}, {G{\'o}mez}, {Besla},
  {Erkal}  \& {Ma}}{{Pe{\~n}arrubia} et~al.}{2016}]{Penarrubia2016}
{Pe{\~n}arrubia} J.,  {G{\'o}mez} F.~A.,  {Besla} G.,  {Erkal} D.,   {Ma}
  Y.-Z.,  2016, \mn@doi [\mnras] {10.1093/mnrasl/slv160}, \href
  {http://adsabs.harvard.edu/abs/2016MNRAS.456L..54P} {456, L54}

\bibitem[\protect\citeauthoryear{{Piatek}, {Pryor}  \& {Olszewski}}{{Piatek}
  et~al.}{2016}]{Piatek2016}
{Piatek} S.,  {Pryor} C.,   {Olszewski} E.~W.,  2016, \mn@doi [\aj]
  {10.3847/0004-6256/152/6/166}, \href
  {http://adsabs.harvard.edu/abs/2016AJ....152..166P} {152, 166}

\bibitem[\protect\citeauthoryear{{Planck Collaboration XVI}}{{Planck
  Collaboration XVI}}{2014}]{Planck2014}
{Planck Collaboration XVI} 2014, \mn@doi [\aap] {10.1051/0004-6361/201321591},
  \href {http://adsabs.harvard.edu/abs/2014A%26A...571A..16P} {571, A16}

\bibitem[\protect\citeauthoryear{{Pontzen}, {Read}, {Teyssier}, {Governato},
  {Gualandris}, {Roth}  \& {Devriendt}}{{Pontzen} et~al.}{2015}]{Pontzen2015}
{Pontzen} A.,  {Read} J.~I.,  {Teyssier} R.,  {Governato} F.,  {Gualandris} A.,
   {Roth} N.,   {Devriendt} J.,  2015, \mn@doi [\mnras]
  {10.1093/mnras/stv1032}, \href
  {http://adsabs.harvard.edu/abs/2015MNRAS.451.1366P} {451, 1366}

\bibitem[\protect\citeauthoryear{{Schaye} et~al.,}{{Schaye}
  et~al.}{2015}]{Schaye2015}
{Schaye} J.,  et~al., 2015, \mn@doi [\mnras] {10.1093/mnras/stu2058}, \href
  {http://adsabs.harvard.edu/abs/2015MNRAS.446..521S} {446, 521}

\bibitem[\protect\citeauthoryear{{Shao}, {Cautun}, {Frenk}, {Gao}, {Crain},
  {Schaller}, {Schaye}  \& {Theuns}}{{Shao} et~al.}{2016}]{Shao2016}
{Shao} S.,  {Cautun} M.,  {Frenk} C.~S.,  {Gao} L.,  {Crain} R.~A.,  {Schaller}
  M.,  {Schaye} J.,   {Theuns} T.,  2016, \mn@doi [\mnras]
  {10.1093/mnras/stw1247}, \href
  {http://adsabs.harvard.edu/abs/2016MNRAS.460.3772S} {460, 3772}

\bibitem[\protect\citeauthoryear{{Shao}, {Cautun}, {Frenk}, {Grand},
  {G{\'o}mez}, {Marinacci}  \& {Simpson}}{{Shao} et~al.}{2018a}]{Shao2018}
{Shao} S.,  {Cautun} M.,  {Frenk} C.~S.,  {Grand} R.~J.~J.,  {G{\'o}mez} F.~A.,
   {Marinacci} F.,   {Simpson} C.~M.,  2018a, \mn@doi [\mnras]
  {10.1093/mnras/sty343}, \href
  {http://adsabs.harvard.edu/abs/2018MNRAS.476.1796S} {476, 1796}

\bibitem[\protect\citeauthoryear{{Shao}, {Cautun}, {Deason}, {Frenk}  \&
  {Theuns}}{{Shao} et~al.}{2018b}]{Shao2018b}
{Shao} S.,  {Cautun} M.,  {Deason} A.~J.,  {Frenk} C.~S.,   {Theuns} T.,
  2018b, \mn@doi [\mnras] {10.1093/mnras/sty1470}, \href
  {http://adsabs.harvard.edu/abs/2018MNRAS.479..284S} {479, 284}

\bibitem[\protect\citeauthoryear{{Shao}, {Cautun}, {Deason}  \& {Frenk}}{{Shao}
  et~al.}{2019}]{Shao2019b}
{Shao} S.,  {Cautun} M.,  {Deason} A.~J.,   {Frenk} C.~S.,  2019, in prep.

\bibitem[\protect\citeauthoryear{{Shaya} \& {Tully}}{{Shaya} \&
  {Tully}}{2013}]{Shaya2013}
{Shaya} E.~J.,  {Tully} R.~B.,  2013, \mn@doi [\mnras] {10.1093/mnras/stt1714},
  \href {http://adsabs.harvard.edu/abs/2013MNRAS.436.2096S} {436, 2096}

\bibitem[\protect\citeauthoryear{{Shi}, {Wang}  \& {Mo}}{{Shi}
  et~al.}{2015}]{Shi2015}
{Shi} J.,  {Wang} H.,   {Mo} H.~J.,  2015, \mn@doi [\apj]
  {10.1088/0004-637X/807/1/37}, \href
  {http://adsabs.harvard.edu/abs/2015ApJ...807...37S} {807, 37}

\bibitem[\protect\citeauthoryear{{Smith}, {Duc}, {Bournaud}  \& {Yi}}{{Smith}
  et~al.}{2016}]{Smith2016}
{Smith} R.,  {Duc} P.~A.,  {Bournaud} F.,   {Yi} S.~K.,  2016, \mn@doi [\apj]
  {10.3847/0004-637X/818/1/11}, \href
  {http://adsabs.harvard.edu/abs/2016ApJ...818...11S} {818, 11}

\bibitem[\protect\citeauthoryear{{Sohn}, {Besla}, {van der Marel},
  {Boylan-Kolchin}, {Majewski}  \& {Bullock}}{{Sohn} et~al.}{2013}]{Sohn2013}
{Sohn} S.~T.,  {Besla} G.,  {van der Marel} R.~P.,  {Boylan-Kolchin} M.,
  {Majewski} S.~R.,   {Bullock} J.~S.,  2013, \mn@doi [\apj]
  {10.1088/0004-637X/768/2/139}, \href
  {http://adsabs.harvard.edu/abs/2013ApJ...768..139S} {768, 139}

\bibitem[\protect\citeauthoryear{{Springel}, {Yoshida}  \& {White}}{{Springel}
  et~al.}{2001}]{Springel2001}
{Springel} V.,  {Yoshida} N.,   {White} S.~D.~M.,  2001, \mn@doi [\na]
  {10.1016/S1384-1076(01)00042-2}, \href
  {http://adsabs.harvard.edu/abs/2001NewA....6...79S} {6, 79}

\bibitem[\protect\citeauthoryear{{Tully}, {Libeskind}, {Karachentsev},
  {Karachentseva}, {Rizzi}  \& {Shaya}}{{Tully} et~al.}{2015}]{Tully2015}
{Tully} R.~B.,  {Libeskind} N.~I.,  {Karachentsev} I.~D.,  {Karachentseva}
  V.~E.,  {Rizzi} L.,   {Shaya} E.~J.,  2015, \mn@doi [\apjl]
  {10.1088/2041-8205/802/2/L25}, \href
  {http://adsabs.harvard.edu/abs/2015ApJ...802L..25T} {802, L25}

\bibitem[\protect\citeauthoryear{{Wang}, {Frenk}  \& {Cooper}}{{Wang}
  et~al.}{2013}]{Wang2013}
{Wang} J.,  {Frenk} C.~S.,   {Cooper} A.~P.,  2013, \mn@doi [\mnras]
  {10.1093/mnras/sts442}, \href
  {http://adsabs.harvard.edu/abs/2013MNRAS.429.1502W} {429, 1502}

\bibitem[\protect\citeauthoryear{{Wang}, {Lin}, {Kang}, {Dutton}, {Yu}  \&
  {Macci{\`o}}}{{Wang} et~al.}{2014}]{Wang2014}
{Wang} Y.~O.,  {Lin} W.~P.,  {Kang} X.,  {Dutton} A.,  {Yu} Y.,   {Macci{\`o}}
  A.~V.,  2014, \mn@doi [\apj] {10.1088/0004-637X/786/1/8}, \href
  {http://adsabs.harvard.edu/abs/2014ApJ...786....8W} {786, 8}

\bibitem[\protect\citeauthoryear{{Yang}, {van den Bosch}, {Mo}, {Mao}, {Kang},
  {Weinmann}, {Guo}  \& {Jing}}{{Yang} et~al.}{2006}]{Yang2006}
{Yang} X.,  {van den Bosch} F.~C.,  {Mo} H.~J.,  {Mao} S.,  {Kang} X.,
  {Weinmann} S.~M.,  {Guo} Y.,   {Jing} Y.~P.,  2006, \mn@doi [\mnras]
  {10.1111/j.1365-2966.2006.10373.x}, \href
  {http://adsabs.harvard.edu/abs/2006MNRAS.369.1293Y} {369, 1293}

\bibitem[\protect\citeauthoryear{{Zel'dovich}}{{Zel'dovich}}{1970}]{Zeldovich1970}
{Zel'dovich} Y.~B.,  1970, \aap, \href
  {http://adsabs.harvard.edu/abs/1970A%26A.....5...84Z} {5, 84}

\bibitem[\protect\citeauthoryear{{Zentner}, {Kravtsov}, {Gnedin}  \&
  {Klypin}}{{Zentner} et~al.}{2005}]{Zentner2005b}
{Zentner} A.~R.,  {Kravtsov} A.~V.,  {Gnedin} O.~Y.,   {Klypin} A.~A.,  2005,
  \mn@doi [\apj] {10.1086/431355}, \href
  {http://adsabs.harvard.edu/abs/2005ApJ...629..219Z} {629, 219}

\makeatother
\end{thebibliography}
\label{lastpage}
\end{document}